\documentclass{aa}  

\usepackage{graphicx}
\usepackage{txfonts}
\usepackage{algorithm,algpseudocode}
\usepackage{subcaption}
\usepackage{comment}
\DeclareMathOperator*{\argmin}{arg\,min}
\usepackage{xcolor}
\usepackage{enumitem}
\usepackage{amsmath}
\usepackage{hyperref}

\newcommand{\algorithmicbreak}{\textbf{break}}

\begin{document}

   \title{An Alternating Minimization Algorithm with Trajectory for \\ Direct Exoplanet Detection}

   \subtitle{The AMAT Algorithm}

   \author{H. Daglayan\inst{1,2}, S. Vary\inst{3}, O. Absil\inst{4}\fnmsep\thanks{F.R.S.-FNRS Senior Research Associate},  F. Cantalloube\inst{5},
   V. Christiaens\inst{4}, N. Gillis\inst{6},  L. Jacques\inst{1}, V. Leplat\inst{7}, \\
          \and
          P.-A. Absil\inst{1}
          }
    \authorrunning{H. Daglayan et al.}

   \institute{ICTEAM Institute, UCLouvain,
              1348 Louvain-la-Neuve, Belgium\\
              \email{hazan.daglayan@uclouvain.be}
         \and
         Vlaamse Instelling voor Technologisch Onderzoek (VITO), 2400 Mol, Belgium 
         \and
         Department of Statistics, University of Oxford, Oxford, United Kingdom 
         \and 
         STAR Institute, Universit\'e de Li\`ege,  4000 Li\`ege, Belgium 
         \and
         Univ. Grenoble Alpes, CNRS, IPAG, F-38000 Grenoble, France
         \and
         Dept.\ of Mathematics and Operational Research, University of Mons, Mons, Belgium 
         \and Innopolis University, Innopolis, Russia 
             }

   \date{Received June 25, 2024; accepted XXX}

  \abstract
   {Effective image post-processing algorithms are vital for the successful direct imaging of exoplanets. Standard PSF subtraction methods use techniques based on a low-rank approximation to separate the rotating planet signal from the quasi-static speckles, and rely on signal-to-noise ratio maps to detect the planet. These steps do not interact or feed each other, leading to potential limitations in the accuracy and efficiency of exoplanet detection.}
   {We aim to develop a novel approach that iteratively finds the flux of the planet and the low-rank approximation of quasi-static signals, in an attempt to improve upon current PSF subtraction techniques.}
   {In this study, we extend the standard L2 norm minimization paradigm to an L1 norm minimization framework to better account for noise statistics in the high contrast images. Then, we propose a new method, referred to as Alternating Minimization Algorithm with Trajectory (AMAT), that makes a more advanced use of estimating the low-rank approximation of the speckle field and the planet flux by alternating between them and utilizing both L1 and L2 norms. For the L1 norm minimization, we propose using L1 norm low-rank approximation (L1-LRA), a low-rank approximation computed using an exact block-cyclic coordinate descent method, while we use randomized singular value decomposition for the L2 norm minimization. Additionally, we enhance the visibility of the planet signal using a likelihood ratio as a postprocessing step.}
   {Numerical experiments performed on a VLT/SPHERE-IRDIS dataset show the potential of AMAT to improve upon the existing approaches in terms of 
   higher S/N, sensitivity limits (contrast curves), and receiver operating characteristic (ROC) curves. Moreover, for a systematic comparison, we used datasets from the exoplanet data challenge to compare our algorithm to other algorithms in the challenge, and AMAT with likelihood ratio map performs better than most algorithms tested on the exoplanet data challenge.}
   {}


   \maketitle
   
%

\section{Introduction} \label{sec:intro}

High-contrast imaging (HCI) is an essential observing technique for exoplanet discovery, especially because of its ability to directly observe young, massive planets orbiting their host stars at large distances~\citep{galicher2023imaging}. This direct observation method is complementary to indirect techniques and also provides crucial information such as bolometric luminosity, effective temperature, surface gravity, composition of the planets and, assuming a formation and evolutionary model, an estimate of their mass~\citep{bowler2016imaging}.
Progress in this field is supported by remarkable technological advances in extreme adaptive optics (AO) systems~\citep{guyon2018extreme} and coronagraphy, for better atmospheric turbulence correction and raw stellar light suppression, respectively. These technologies equip new generation HCI instruments, such as VLT/SPHERE~\citep{beuzit2019sphere}, Gemini/GPI~\citep{macintosh2014first} or Subaru/SCExAO~\citep{martinache2009subaru} allowing them to achieve unprecedented raw contrasts. Despite these important steps, direct imaging of exoplanets remains a challenging task, and only 1\% of known exoplanets have been discovered using this method \citep{nasa}. This limitation underscores the inherent difficulty in distinguishing these faint planets from the overwhelming brightness of their host stars. Star light residuals under the form of speckles, caused by atmospheric turbulence and imperfections in telescopes and instruments, are very similar in shape and contrast to planets, posing a major challenge to direct imaging. To address this, angular differential imaging (ADI) has emerged as a common strategy~\citep{marois2006angular}. ADI involves capturing a sequence of frames in pupil-stabilized mode, wherein the telescope tracks the star's motion over time, keeping it centered in the image. This approach results in the star and speckles appearing static or quasi-static, while planets exhibit movement as a function of the parallactic angle due to Earth's rotation. By exploiting this differential motion, ADI enables the isolation and detection of exoplanetary signals from the surrounding noise, increasing the chances of successful direct imaging.

Using ADI sequences, several post-processing methods have been proposed. Among the most common ones are those that build a model for the stellar PSF (including the static and quasi-static speckle field), and subtract it to detect planets. To build a PSF model, methods such as principal component analysis \citep[PCA,][]{Amara2012pynpoint, Soummer2012Detection} or nonnegative matrix factorization \citep[NMF,][]{GomezGonzalez2017VIP,ren2018non} aim to obtain a low-rank approximation of the time-by-pixel matrix containing the ADI observing sequence. After subtracting the PSF model from each frame, the residual matrix consists of both planetary signals and noise, which includes some residual stellar signal. 
Some studies suggest modeling the residual matrix as a sparse matrix and noise \citep[LLSG and LRPT,][]{GomezGonzalez2016Lowranka,vary2023low}. The Locally Optimized Combination of Images \citep[LOCI,][]{lafreniere2007new} employs a least-squares approach to construct a model PSF, with variants such as TLOCI \citep{marois2010exoplanet} and MLOCI \citep{wahhaj2015improving}. Once the PSF model is obtained and the residual matrix is derived, one must still use a method to extract the planetary signal. The most popular way to do so is to build a signal to noise ratio (S/N) map \citep{mawet2014fundamental}, an alternative being the use of a standardized trajectory intensity mean (STIM) map \citep{pairet2019stim}. More recent methods to extract the planetary signal from residual data cubes include the regime-switching model \citep[RSM,][]{dahlqvist2020regime}, which attempts planetary detection using multiple PSF subtraction techniques at the same time, and the likelihood ratio map \citep[LRM,][]{daglayan2022likelihood}, which proposes a map consisting of likelihood ratios based on maximum likelihood estimation. Besides PSF model subtraction, other post-processing methods are based on inverse problem approaches to estimate the speckle field and planetary signal simultaneously in a maximum-likelihood approach, such as ANDROMEDA \citep{cantalloube2015direct}, FMMF \citep{ruffio2017improving}, PACO \citep{flasseur2018exoplanet}, and SNAP \citep{thompson2021improved}. Additionally, another type of algorithms utilizes machine learning for planet detection, such as SODIRF and SODINN \citep{gonzalez2018supervised}, and their refined variant NA-SODINN \citep{cantero2023sodinn}. As an aside, most of these post-processing methods are designed or optimized to detect point sources around the target star, and generally struggle to reconstruct extended sources like circumstellar disks. Iterative versions of PSF-subtraction methods, including iterative PCA, have recently been proposed to address this shortcoming \citep{pairet2021mayonnaise,refId0, juillard2023inverse}.

One recurrent assumption made by the post-processing methods described above is that residual noise after PSF subtraction is Gaussian. However, recent studies  \citep{ruffio2017improving, pairet2019stim, dahlqvist2020regime, daglayan2022likelihood, cantero2023sodinn} have shown that the noise in the residual cube and/or processed frame, particularly in the tails of the distribution, tends to be non-Gaussian. In light of this, in Sect.~\ref{sec:amat}, our paper proposes a low-rank approximation based on the Laplacian distribution, termed L1-LRA, which leverages the L1 norm. We demonstrate that this approach more accurately fits the data. On the other hand,  the minimization of the L2 norm is computationally easier due to the smoothness of the objective function. Subsequently, we present an iterative method named Alternating Minimization with Trajectory (AMAT), designed to enhance algorithmic performance and more effectively differentiate between planetary signals and static or quasi-static signals using both L1 and L2 norm. Additionally, we establish that this iterative method, in its L2 norm version, slightly outperforms state-of-the-art methods for determining the planetary flux. To assess the performance of our proposed algorithms, we employ various benchmarks, such as S/N maps, contrast curves, and Receiver Operating Characteristic (ROC) curves in Sect.~\ref{sec:results}. For these empirical analyses, we use an ADI cube obtained on 51 Eri with the VLT/SPHERE-IRDIS instrument as used in the \cite{samland_spectral_2017} publication, containing 256 frames obtained in the K1 (2.11 {$\mu$m}) band over a parallactic angle range of 42$^{\circ}$ \citep{samland_spectral_2017}. Additionally, we leverage datasets from the exoplanet data challenge for comparative analysis against other state-of-the-art methods~\citep{cantalloube2020exoplanet}. In Sect.~\ref{sec:likelihood}, we propose to further improve the performance of our method by computing an LRM based on the residual cube processed with the AMAT algorithm, instead of using a standard S/N map. Section~\ref{sec:conclusion} concludes
the paper. 

Preliminary results related to the present work appeared in the proceedings of two machine learning conferences~\citep{daglayan2023esann,daglayan2023l1lra}. \citet{daglayan2023l1lra} suggested the potential use of L1 norm low-rank approximation for exoplanet detection, while \citet{daglayan2023esann} provided a brief description of the AMAT algorithm and performed ablation studies to evaluate its performance. Both studies evaluated the algorithms using a single dataset. This paper expands upon these descriptions, offering a comprehensive explanation of the algorithms and comparing them using various metrics. We demonstrate the performance of the algorithm across different datasets to illustrate its data-independent capabilities. Additionally, we explore the application of the AMAT algorithm for flux estimation and compare its efficacy with existing methods.

\section{The AMAT algorithm}\label{sec:amat}

In this section, we present a novel method that employs an iterative technique for exoplanet detection that distinguishes planetary signals from the star and its associated speckles, as well as from the sky background. This method aims to find a low-rank matrix that better fits the static/quasi-static signal and reveals the planetary signal more clearly, and proposes the use of the L1 norm as a solution to mitigate the effects of Laplacian noise. Firstly, we describe the L1 low-rank approximation (L1-LRA) in detail, explaining its reliance on the L1 norm. We then present the AMAT algorithm, which is designed to accommodate both L1 and L2 norm scenarios.

\subsection{L1 low-rank approximation}\label{sec:l1lra}

\begin{figure}[!t]
    \centering
    \begin{subfigure}[b]
    {\textwidth}
        \includegraphics[width=0.4\textwidth]{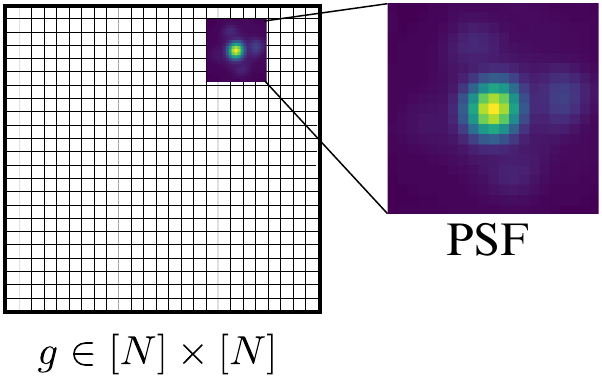}
    \end{subfigure}
    \begin{subfigure}[b]
    {\textwidth}
        \includegraphics[width=0.4\textwidth]{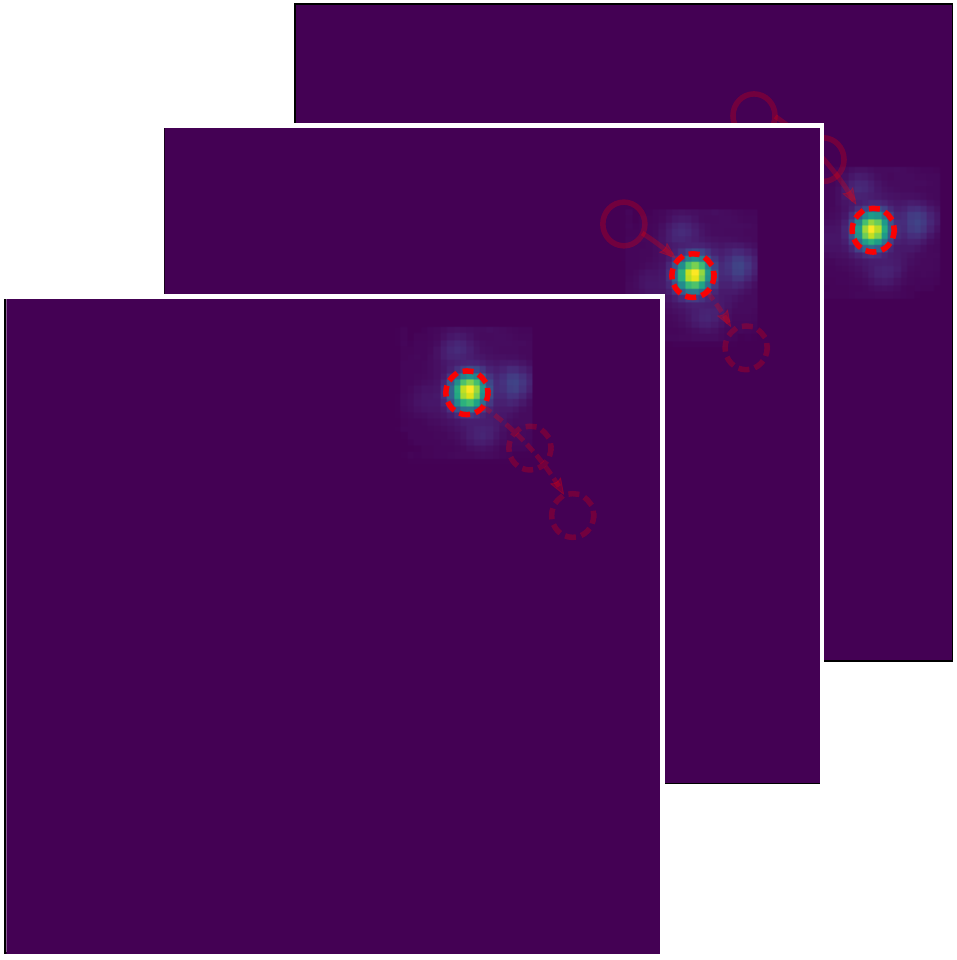}
    \end{subfigure}
    \caption{The cube of the planet signature constructed by rotating the position $g$ of the PSF function along the trajectory.}\label{fig:pg}
\end{figure}

Let $M$ be a matrix in $\mathbb{R}^{t\times n^2 }$ containing observations of $t$ unfolded frames, where each row represents a single vectorized frame of size $n \times n$. Assuming there is a single planet located at position $g\in[n]\times[n]$ within the first frame, with $[n]=\{1,2,\dots,n\}$, the model for $M$ can be written as
\begin{equation}\label{eq:data_model}
    M = L + a_gP_g + E, \quad \mathrm{rank}(L)\leq k, \quad P_g \in \mathcal{P},
\end{equation}
where $L$ denotes the low-rank model for the stellar diffraction pattern, $E$ stands for the noise, $a_g$ is the intensity of the planet referred to as the \emph{flux}, $P_g \in \mathcal{P} \subset \mathbb{R}^{t\times n^2}$ is the planet signature along the trajectory, illustrated in Fig. \ref{fig:pg}, and $\mathcal{P}$ is the set of all feasible planet signatures. To construct $P_g$, we start from a time-by-pixel matrix with zero entries, and in each unfolded frame (\emph{i.e.}, row of the matrix), we place a copy of the normalized reference PSF at the location occupied by a planet that is at position $g$ in the first frame. We normalize the reference PSF using the method described in VIP \citep{GomezGonzalez2017VIP}. This involves dividing the pixel values by the sum of pixel intensities measured within a full-width half maximum (FWHM) aperture. As a result, 
$P_g$ is a constant matrix for one single planet position~$g$.

In models employing low-rank approximations, selecting the appropriate rank value is critical. If the rank is too small, speckle signals may persist in the residual $M-L$, making it challenging to distinguish the signal of the planet from the speckles. In contrast, the signal of the planet may be absorbed by the low-rank matrix if the rank is too large, making it more difficult to locate the planet signal in the residual.

When the error $E$ follows Gaussian distribution, the maximum likelihood estimator for $L$ is obtained by minimizing the L2 norm. Classical methods like PCA and LLSG fit the low-rank component as follows: 
\begin{align}
    \hat{L} = \underset{L}{\arg\min} \|M - L \|_2 \quad 
    \text{subject to} \quad \mathrm{rank}(L) \leq k, 
    \label{pro:PCA}
\end{align}
where $\|A\|_2$ denotes the entry-wise L2 norm of $A$ (the Frobenius norm), which can be solved using the truncated SVD according to the Eckart–Young–Mirsky theorem~\citep{eckart1936approximation}. Such an approach to estimate $L$ has been used in~\cite{Amara2012pynpoint,Soummer2012Detection, GomezGonzalez2016Lowranka}. Recent findings indicate that the error term $E$ exhibits heavy tails, aligning more closely with the Laplacian distribution \citep{pairet2019stim, cantero2023sodinn}. Consequently, we propose fitting the low-rank component using the component-wise L1 norm
\begin{align}
    \hat{L} = \underset{L}{\arg\min} \|M - L \|_1 \quad \mathrm{subject \ to}\quad \mathrm{rank}(L) \leq k
    \label{pro:l1}.
\end{align}
This approach allows us to maintain a consistent noise assumption across the low-rank speckle subtraction. Moreover, in PCA, a common issue is the sensitivity to outliers when using the L2 norm. In contrast, the L1 norm demonstrates a robust approach to outliers, leading to a better fit~\citep{ke2003robust,ke2005robust,song2017low}. This makes the L1 norm a more suitable choice for data with potential outliers.

The L1 low-rank approximation in Eq.~\eqref{pro:l1} is, however, an NP-hard problem, even in the rank-one case~\citep{gillis2018complexity}. 
Hence, most algorithms to tackle Eq.~\eqref{pro:l1}, such as alternating convex optimization \citep{ke2005alternating}, the Wiberg algorithm \citep{eriksson2010efficient}, and augmented Lagrangian approaches \citep{zheng2012practical}, do not guarantee to find a global optimal solution, unlike in the case of PCA. Moreover, the computed solutions are sensitive to the initialization of the algorithms.  
We use Algorithm~\ref{alg:l1} (L1-LRA) suggested by \cite{gillis2018complexity} to solve Eq.~\eqref{pro:l1}. It solves the problem using an exact block-cyclic coordinate descent method, where the blocks of variables are the columns of $\hat{U}$ and 
$\hat{V}$ of the low-rank approximation $\hat{L}=\hat{U}\hat{V}^\top$. This algorithm relies on the fact that the columns of the matrix $\hat V \in \mathbb{R}^{n^2\times k}$ form a basis of a subspace of $\mathbb{R}^{n^2}$ that best approximates the data frames in the L1 sense and, for each data frame, the corresponding row of $\hat{U} \in \mathbb{R}^{t\times k}$ contains the weights of the L1-best approximation of the data frame in this basis. 
Here, $A_j$ denotes $j$-th column of the matrix $A$. 

\begin{algorithm}
    \caption{L1-LRA \citep{gillis2018complexity}\label{alg:l1} }
    \hspace*{\algorithmicindent} \textbf{Input:} Image sequence $M \in \mathbb{R}^{t\times n^2}$, the initial components 
    \hspace*{\algorithmicindent} $\hat{U}\in \mathbb{R}^{t\times k}$ and $\hat{V} \in \mathbb{R}^{n^2\times k}$  of $M$ (default initialization with the 
    \hspace*{\algorithmicindent} randomized SVD), rank $k$, maximum number of iteration \hspace*{\algorithmicindent} $\ell_\textrm{max}$. \\
    \hspace*{\algorithmicindent} \textbf{Output: }the components $\hat{U}$ and $\hat{V}$.

    \begin{algorithmic}[1]
    \For {$\ell = 1: \ell_\mathrm{max}$} 
        \State $R = M-\hat{U}\hat{V}^\top$ 
        \For {$j=1:k$} 
        \State$R \leftarrow R+\hat{U}_j\hat{V}_j^\top$ 
        \State $\hat{U}_j \leftarrow \underset{{u \in \mathbb{R}^t}}{\argmin}\|R-u \hat{V}_j^\top\|_1$  
        \State $\hat{V}_j \leftarrow  \underset{{v \in \mathbb{R}^{n^2}}}{\argmin} \| R^\top-v \hat{U}_j^\top\|_1$ 
        \State $R \leftarrow R-\hat{U}_j\hat{V}_j^\top$ 
        \EndFor
    \EndFor
    \end{algorithmic}
\end{algorithm}

To solve the minimization problem in steps 5-6 of Algorithm~\ref{alg:l1}, we use the exact method from~\cite{gillis2011dimensionality}; these subproblems are weighted median problems that can be solved in closed form. In our experiments, we apply an annular version, similar to annular PCA \citep[AnnPCA,][]{Absil2013,GomezGonzalez2017VIP}, that selects only the pixels of $M$ in a certain annulus. Indeed, as the intensities of pixels decrease away from the star, it is usually better to calculate the low-rank approximation of each annulus separately.

\begin{figure*}[!t]
    \centering
    \begin{subfigure}[b]{0.48\textwidth}
        \centering
        \includegraphics[width=\textwidth]{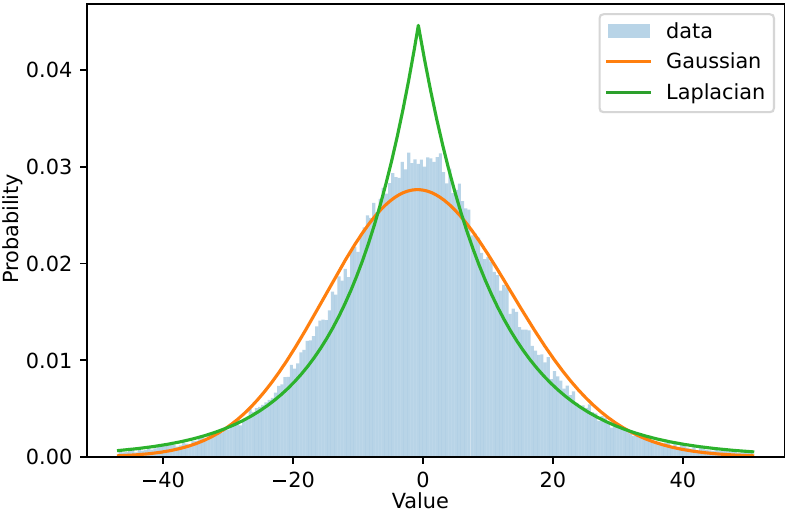}
        \caption{PCA at 4$\lambda/D$ \label{fig:pca_small}}
    \end{subfigure}
    \begin{subfigure}[b]{0.48\textwidth}
        \centering
        \includegraphics[width=\textwidth]{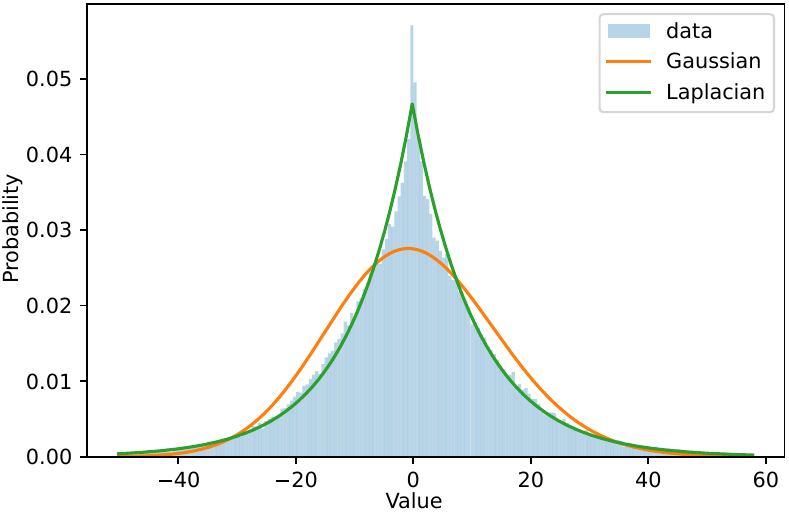}
        \caption{L1-LRA at 4$\lambda/D$ \label{fig:l1_small}}
    \end{subfigure}
    \begin{subfigure}[b]{0.48\textwidth}
        \centering
        \includegraphics[width=\textwidth]{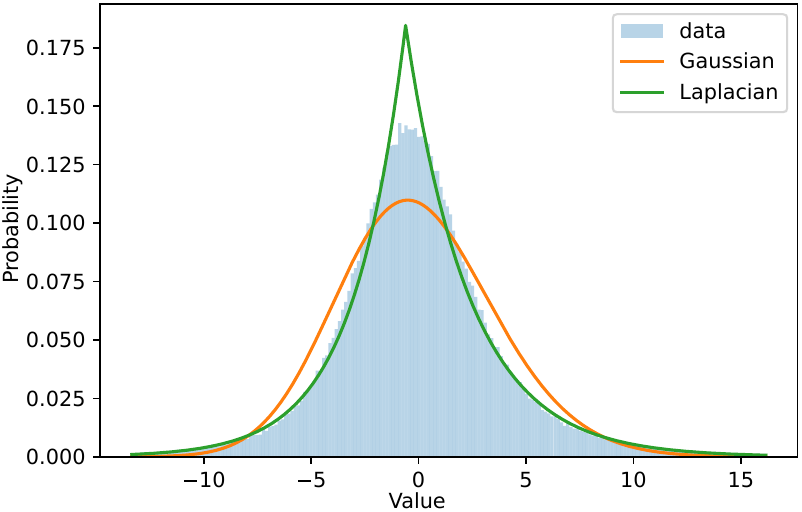}
        \caption{PCA at 8$\lambda/D$ \label{fig:pca_large}}
    \end{subfigure}
    \begin{subfigure}[b]{0.48\textwidth}
        \centering
        \includegraphics[width=\textwidth]{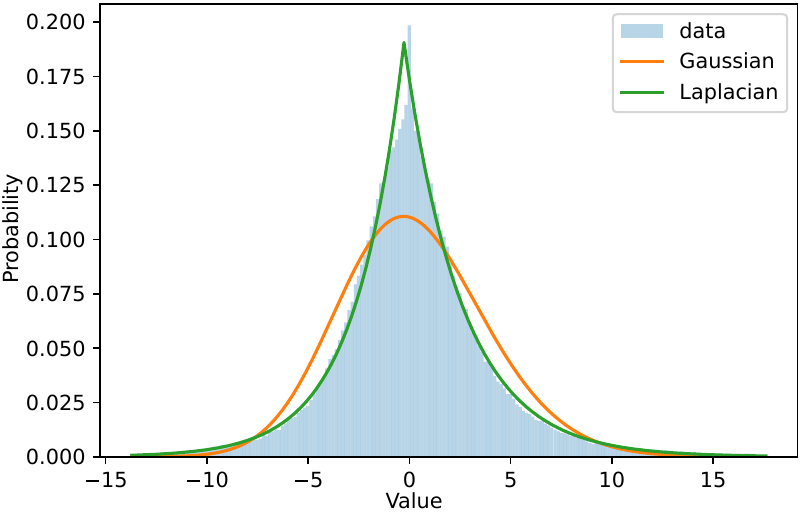}
        \caption{L1-LRA at 8$\lambda/D$ \label{fig:l1_large}}
    \end{subfigure}

       \caption{Histograms of residual cube after low-rank approximation applied for small and large separations for a 51 Eri dataset } \label{fig:distribution}
\end{figure*}

In order to analyze the suitability of different noise assumptions, we fit Gaussian and Laplacian distributions to the residual data, \emph{i.e.}, the data after subtracting the low-rank component using PCA or L1-LRA. We look at two different annuli separately, one that is close to the star at $4\lambda/D$ separation and one more distant from the star at $8\lambda /D$, and measure the goodness of fit visually. In Fig.~\ref{fig:distribution}, we observe that the residual data follows somewhere between Gaussian and Laplacian on the peak and Laplacian on the tails of the distribution after applying PCA. However, after applying L1-LRA, the Laplace distribution provides a better fit for the residual cube distribution in general for both small and large separations. This supports that L1 is the indicated norm in a noise model where the error follows a Laplacian distribution~\citep{gao2009robust}.

\subsection{The AMAT Algorithm: Planet detection}
Instead of traditional approaches of estimating first the low-rank component $L$ and then by estimating the flux $a_g$, we propose to estimate them simultaneously in the following optimization problem
\begin{equation}\label{eq:main_problem}
    \min_{L \in\mathbb{R}^{t \times n^2}, a_g \in \mathbb{R}} \left\| M - L - a_g P_g \right\|_{\#} \quad \mathrm{s.t.}  \quad \mathrm{rank}(L)\leq k, \quad P_g \in \mathcal{P},
\end{equation}
where $\| \cdot \|_{\#}$ denotes either the L1 or L2 norm depending on the assumed distribution of the error $E$ in Eq.~\eqref{eq:data_model}. The optimization problem Eq.~\eqref{eq:main_problem} is addressed by alternatingly solving the following two sub-problems until a stopping criterion is met:
\begin{subequations}
\begin{equation}
    L^{(i)}  = \argmin_{L \in \mathbb{R}^{t \times n^2}} \| M - L - a_g^{(i-1)} P_g \|_{\#}, \label{eq:L} 
\end{equation}
\begin{equation}
    a_g^{(i)}  = \argmin_{a \in \mathbb{R}} \| M - L^{(i)} - a P_g \|_{\#} \label{eq:a}.
\end{equation}
\end{subequations}
The stopping criterion is defined as either reaching a maximum number of iterations or ensuring that the relative changes in the intensity $a_g^{(i)}$ are less than a specified threshold. 

When we select the norm L2 in~\eqref{eq:main_problem} as given in Algorithm \ref{alg:amat_l2}, computing $L^{(i)}$ in Eq.~\eqref{eq:L} amounts to a $k$-truncated SVD, denoted by $\mathrm{H}^\mathrm{SVD}_k({\cdot})$. In practice, in order to speed up the computations, we compute $L^{(i)}$ by a randomized SVD of $M-a_g^{(i-1)}P_g$~\citep{halko2011finding}. We have checked that the resulting approximation does not affect the planet detection performance. The optimal value of $a_g^{(i)}$ can be computed by cross correlating the residual cube $R_g = M-L^{(i)}$ with the planet signature $P_g$: 
\begin{equation}\label{eq:a_g_gauss}
    a^{(i)}_g = \frac{\sum_{(\theta, r) \in \Omega_g}{R_g(\theta,r)P_g(\theta,r)\sigma^{-2}_{R_g(r)}}}
    {\sum_{(\theta, r) \in \Omega_g}{(P_g(\theta,r))^2 \sigma^{-2}_{R_g(r)}}},
\end{equation}
where 
$\sigma_{R}^2$ is the empirical variance of the residual frames computed along the time dimension and $\Omega_g$ is the set of indices $(\theta, r)$ of pixels whose distance from the trajectory is smaller than half the chosen aperture diameter $\rho\lambda/D$, with $\rho>0$
\begin{equation} 
\label{eq:listind}
	\Omega_g = \left\{
		(\theta, r) \in [t] \!\times\! [n]^2
		\,\big\rvert\,
		\|r - g_t\|_2 < \tfrac{1}{2}  \tfrac{\lambda}{D}  \rho 
	\right\}.
\end{equation}

\begin{algorithm}
    \caption{AMAT$_\textrm{L2}$\label{alg:amat_l2} }
    \hspace*{\algorithmicindent} \textbf{Input:} Image sequence $M \in \mathbb{R}^{t\times n^2}$, possible trajectories \hspace*{\algorithmicindent} $\mathcal{P}$, rank $k$, maximum number of iteration for AMAT $i_{\textrm{max}}$, \hspace*{\algorithmicindent} threshold for relative change $\epsilon$ \\
    \hspace*{\algorithmicindent} \textbf{Output: }Low rank component $L^{(i)}$ and flux $a_g^{(i)}$ for each \hspace*{\algorithmicindent} trajectory.
    \begin{algorithmic}[1]
    \For {all trajectories $P_g \in \mathcal{P}$} 
        \State $a_g^{(0)} = 0$
        \For {$i = 1: i_{\textrm{max}}$} 
        \State $\dot{U}, \dot{S}, \dot{V}^\top =$ $\mathrm{H}^\mathrm{SVD}_k(M- a_g^{(i-1)} P_g)$
        \State $L^{(i)} = \dot{U}\dot{S}\dot{V}^\top$
        \State Compute $a^{(i)}_g$ using \eqref{eq:a_g_gauss}
        \If {$|a_g^{(i)}-a_g^{(i-1)}|/|a_g^{(i)}|<\epsilon$}
        \State \algorithmicbreak
        \EndIf
        \EndFor
    \EndFor
    \end{algorithmic}
\end{algorithm} 
If \eqref{eq:main_problem} is set with the norm L1, we solve the problem Eq.~\eqref{eq:L} with the algorithm suggested in Sect.~\ref{sec:l1lra}. We initialize the algorithm with the randomized SVD solution in Algorithm~\ref{alg:amat_l1}. Then, we tackle  the problem Eq.~\eqref{eq:a} by 
\begin{align}\label{eq:a_l1_norm}
        \begin{split}
	a^{(i)}_g
	= \argmin_{a} \sum_{(\theta, r) \in {\Omega}_g}
	\frac{|{R_g(\theta, r) - a P_g(\theta, r)}|}{\sigma_{R(r)}}.
        \end{split}
\end{align}
Solving Eq.~\eqref{eq:a_l1_norm} is an instance of the weighted least absolute deviation (LAD) problem, which, unlike least squares, does not have a closed form solution. In general, L1 minimization can be solved by a number of efficient iterative methods, however, in our specific case, it is possible to compute the solution even more efficiently. Since the objective function is a convex piecewise linear function $\mathbb{R} \rightarrow \mathbb{R}$ with intervals between the points $R(\theta,r)/P_g(\theta,r)$, $(\theta, r) \in {\Omega}_g$, its minimum is attained at one of the $(tn^2)$ kink points. These kink points are the boundary points where different linear segments of the piecewise function meet, and they can be easily searched exhaustively.

\begin{algorithm}
    \caption{AMAT$_\textrm{L1}$\label{alg:amat_l1} }
    \hspace*{\algorithmicindent} \textbf{Input:} Image sequence $M \in \mathbb{R}^{t\times n^2}$, possible trajectories \hspace*{\algorithmicindent} $\mathcal{P}$, rank $k$, maximum number of iteration for AMAT $i_{\textrm{max}}$, \hspace*{\algorithmicindent} maximum number of iteration for L1-LRA $\ell_{\textrm{max}}$, threshold \hspace*{\algorithmicindent} for relative change $\epsilon$ \\
    \hspace*{\algorithmicindent} \textbf{Output: }Low rank component $L^{(i)}$ and flux $a_g^{(i)}$ for each \hspace*{\algorithmicindent} trajectory.
    \begin{algorithmic}[1]
    \For {all trajectories $P_g \in \mathcal{P}$} 
        \State $a_g^{(0)} = 0$ 
        \State $\dot{U}, \dot{S}, \dot{V}^\top =$ $\mathrm{H}^\mathrm{SVD}_k(M- a_g^{(0)} P_g)$
        \State $U^{(0)} = \dot{U}\dot{S}$; $\quad V^{(0)} =\dot{V}$ 
        \For {$i = 1: i_{\textrm{max}}$} 
        \State $U^{(i)}, V^{(i)} = $\\ \hspace*{\algorithmicindent} \hspace*{\algorithmicindent} L1-LRA($M- a_g^{(i-1)} P_g, U^{(i-1)}, V^{(i-1)}, k, \ell_{\textrm{max}}$)
        \State $L^{(i)} = U^{(i)}{V^{(i)}}^\top$ 
        \State Compute $a^{(i)}_g$ using \eqref{eq:a_l1_norm}
        \If {$|a_g^{(i)}-a_g^{(i-1)}|/|a_g^{(i)}|<\epsilon$}
        \State \algorithmicbreak
        \EndIf
        \EndFor
    \EndFor
    \end{algorithmic}
\end{algorithm}

\begin{figure}[!t]
    \centering
    \begin{subfigure}[b]
    {0.40\textwidth}
        \includegraphics[width=\textwidth]{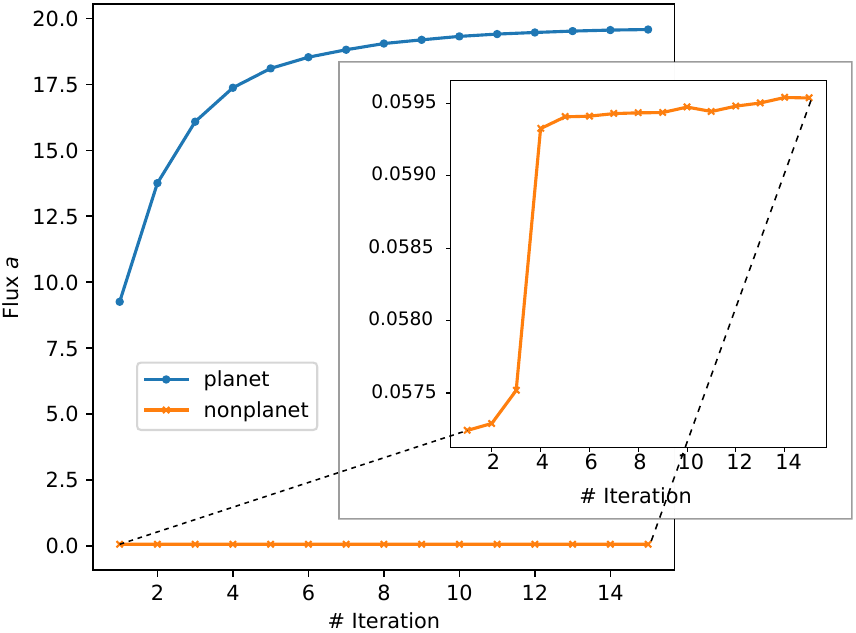}
    \end{subfigure}
    \begin{subfigure}[b]
    {0.40\textwidth}
        \includegraphics[width=\textwidth]{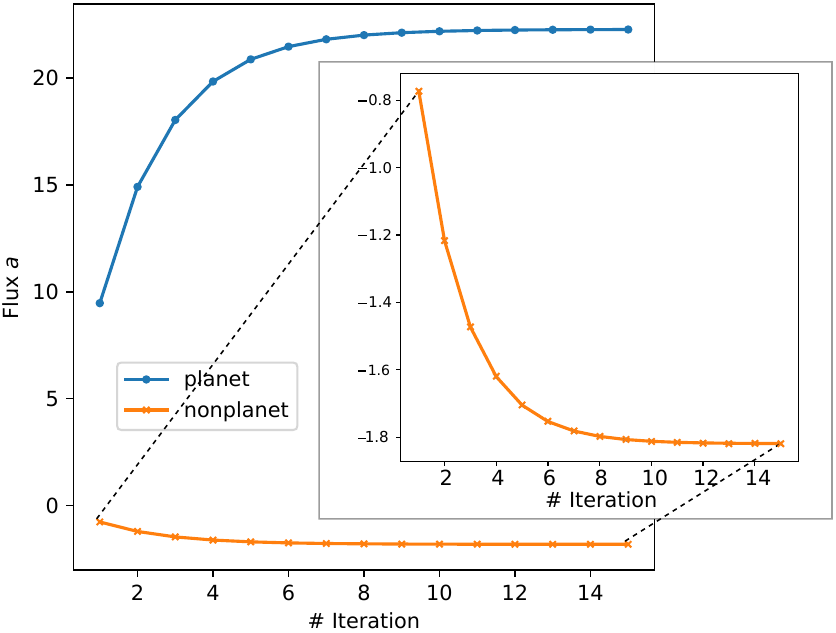}
    \end{subfigure}
\caption{Intensity $a$ of the planet against the number of iterations of Eqs.~\eqref{eq:L} -- ~\eqref{eq:a}. Top: The intensity $a_g^{(i)}$ is obtained using L1 norm. Bottom: The intensity $a_g^{(i)}$ is obtained using L2 (Frobenius) norm. The blue plots show how the intensity changes in each iteration when we choose $P_g$ in the location of the planet. The orange plots show how the intensity changes in each iteration when we choose $P_g$ in a location without a planet.}\label{fig:iteration}
\end{figure}

As part of our evaluation of the performance of the AMAT algorithms, we delve into the impact of iteration counts on its results. Iterative processes are integral to the algorithm, making it essential to investigate how its outcomes evolve over successive iterations. To illustrate the behavior of the algorithm, we show in Fig.~\ref{fig:iteration} how the estimated flux $a_g$ evolves in iterations when the trajectory corresponds to the correct location of the planet and when it does not. We simulate this scenario by injecting a fake planet into the 51 Eri dataset to observe the changes in the flux values at the planet pixels. The results for both norms show that there is a considerable amount of change in the flux $a_g$ for the $P_g$ in the planet pixels, whereas the change in the flux $a_g$ for the $P_g$ in the pixels without a planet is very small.

\begin{figure*}[!t]
    \centering
        \includegraphics[width=\textwidth]{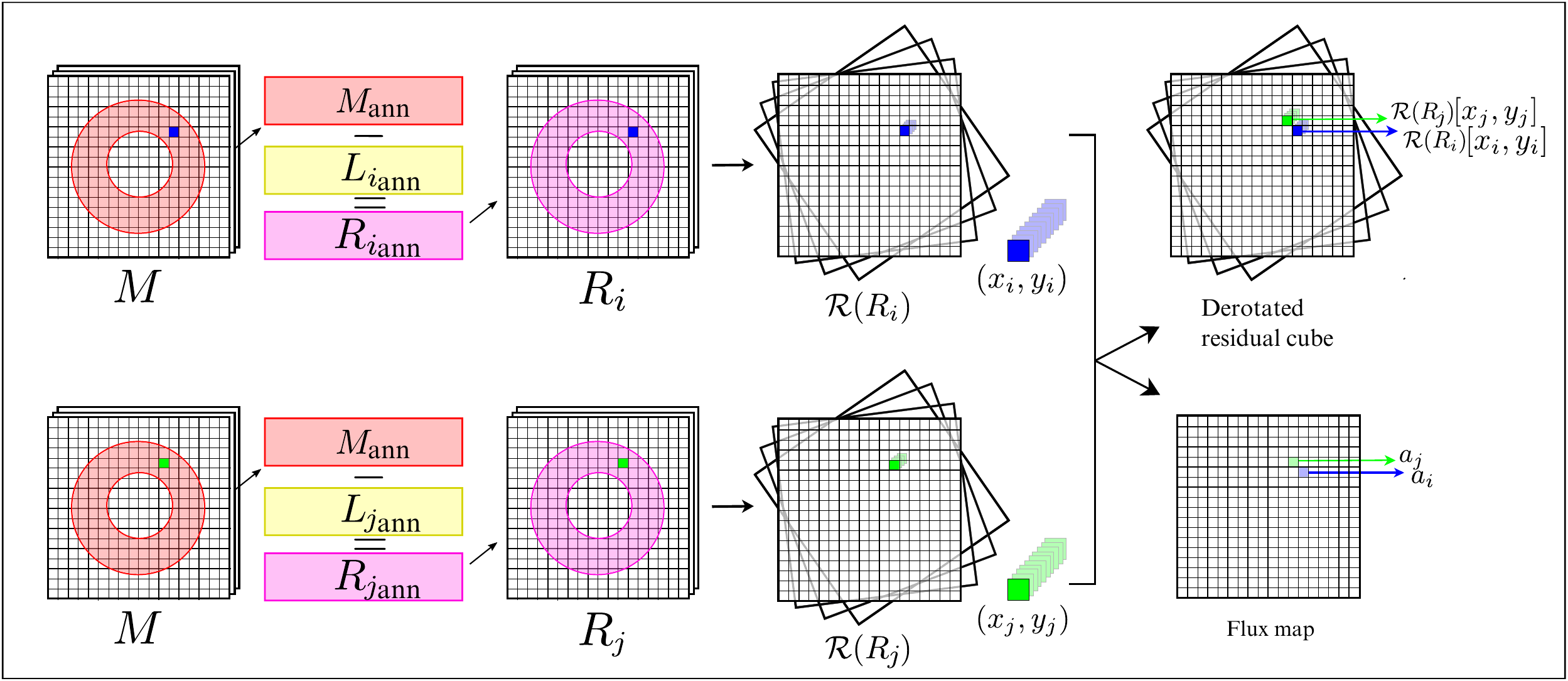}
    \caption{Representation of the residual cube and flux map construction. Example pixels, visualized in blue and green, are selected from possible trajectories. For each selected pixel, a low-rank matrix is obtained, which is used to construct a residual cube and calculate a flux value. After derotating the residual cube, an array is produced. Combining these arrays from all possible trajectories forms the derotated residual cube. Similarly, flux values from all points along the possible trajectories are combined to form the flux map.}\label{fig:combine_cube}
\end{figure*}

We define the set $G$ of positions of the planet as a collection of all points, excluding the pixels of the host star and the pixels in the corners and edges of the images. In our algorithm, we apply the following steps to construct the residual cube and the flux map:
\begin{enumerate}
    \item \label{it:alg_1} Select a pixel, denoted as $(x, y)$ from the set $G$. 
    \item Take the pixels of the annulus centered on the star with an inner radius of $r - $FWHM and an outer radius of $r + $FWHM, where $r$ represents the distance from the center of the star to the point $(x, y)$ resulting in an annulus width of 2FWHM.
    \item \label{it:alg_3} Apply the AMAT algorithm: iteratively estimate a low-rank matrix that encapsulates the background, including quasi-static speckles, and the flux $a_g$ associated with the exoplanet.
    \item For residual cube:
    \begin{enumerate}[ref=\theenumi(\alph*)]
        \item\label{it:alg_4a} Subtract the low rank matrix from the annulus pixels of original data matrix.
        \item\label{it:alg_4b} Assign the residual values from each frame corresponding to $(x, y)$ into the same positions in an empty cube referred to as the derotated residual cube.
        \item Repeat the processes \ref{it:alg_1}-\ref{it:alg_4b} for all pixels in set $G$ to fill the derotated residual cube.
    \end{enumerate}
    
    \item For flux map:
    \begin{enumerate}[ref=\theenumi(\alph*)]
        \item \label{it:alg_5a} Assign the flux $a_g$ to $(x, y)$ into an empty frame ensuring it matches the dimensions of any frame in the data cube.
        \item Repeat the processes \ref{it:alg_1}-\ref{it:alg_3} and \ref{it:alg_5a} for all pixels in set $G$ to fill the flux map.
    \end{enumerate}

\end{enumerate}
For every $g$ in $G$, the flux $a_g$ can be interpreted as follows: if there is a single planet located at $g$ in the first frame (and assuming that Eq.~\eqref{eq:a} is solved exactly), then its flux that best explains the data is $a_g$. Since we apply this algorithm for each position $g$, we calculate $a_g$ for each trajectory, regardless of whether there is more than one planet or no planets at all.
The entire process is illustrated in Fig. \ref{fig:combine_cube} and implemented in the AMAT Python package\footnote{The AMAT algorithm Python package is available on GitHub: \href{https://github.com/hazandaglayan/AMAT}{https://github.com/hazandaglayan/AMAT}.}.
From the flux map, detection is then performed by producing an S/N map and applying a threshold, as described in Sect.~\ref{sec:snr_contrastcurve}. As for the residual cube, it is made use of in Sect.~\ref{sec:likelihood}.

In existing studies that employ iterative PCA for disk and exoplanet detection~\citep{pairet2021mayonnaise, refId0, juillard2023inverse}, the process typically involves applying the full process of PCA steps to generate a median frame. This median frame is then rotated according to parallactic angles and subtracted from each frame of the cube, leading to the creation of a new residual cube and a new median frame. This cycle of rotating, subtracting, and then creating new residual and median frames is the part that is iteratively repeated. In contrast, when using the AMAT algorithm, which is tailored for point source detection, instead of subtracting the median frame, we subtract a matricized cube $P_g$ multiplied by the intensity $a_g$ identified at each step, effectively isolating only the planetary signature because the pixels outside the trajectory are zero. This approach significantly enhances the separation performance between the background signal and the planetary signature, improving overall detection efficiency.

\subsection{Flux estimation}

The detection of exoplanets is followed by the characterization of the planets, which encompasses estimating their positions and the intensity relative to the host star. Different algorithms, such as the negative fake companion (NEGFC) method  ~\citep{lagrange2010giant,marois2010exoplanet,wertz2017vlt}, ANDROMEDA via maximum likelihood estimation~\citep{cantalloube2015direct}, PACO estimation~\citep{flasseur2018exoplanet} are employed for this purpose. 
In our work, when using the AMAT algorithm, we produce a flux map consisting of the intensities $a_g$ corresponding to each trajectory. Following the planet detection, we use this flux map to estimate the intensity of the planet at its detected location. We obtain these intensity values directly from the flux map without accounting for sub-pixel precision.

\begin{figure}[!t]
    \centering
        \includegraphics[width=0.48\textwidth]{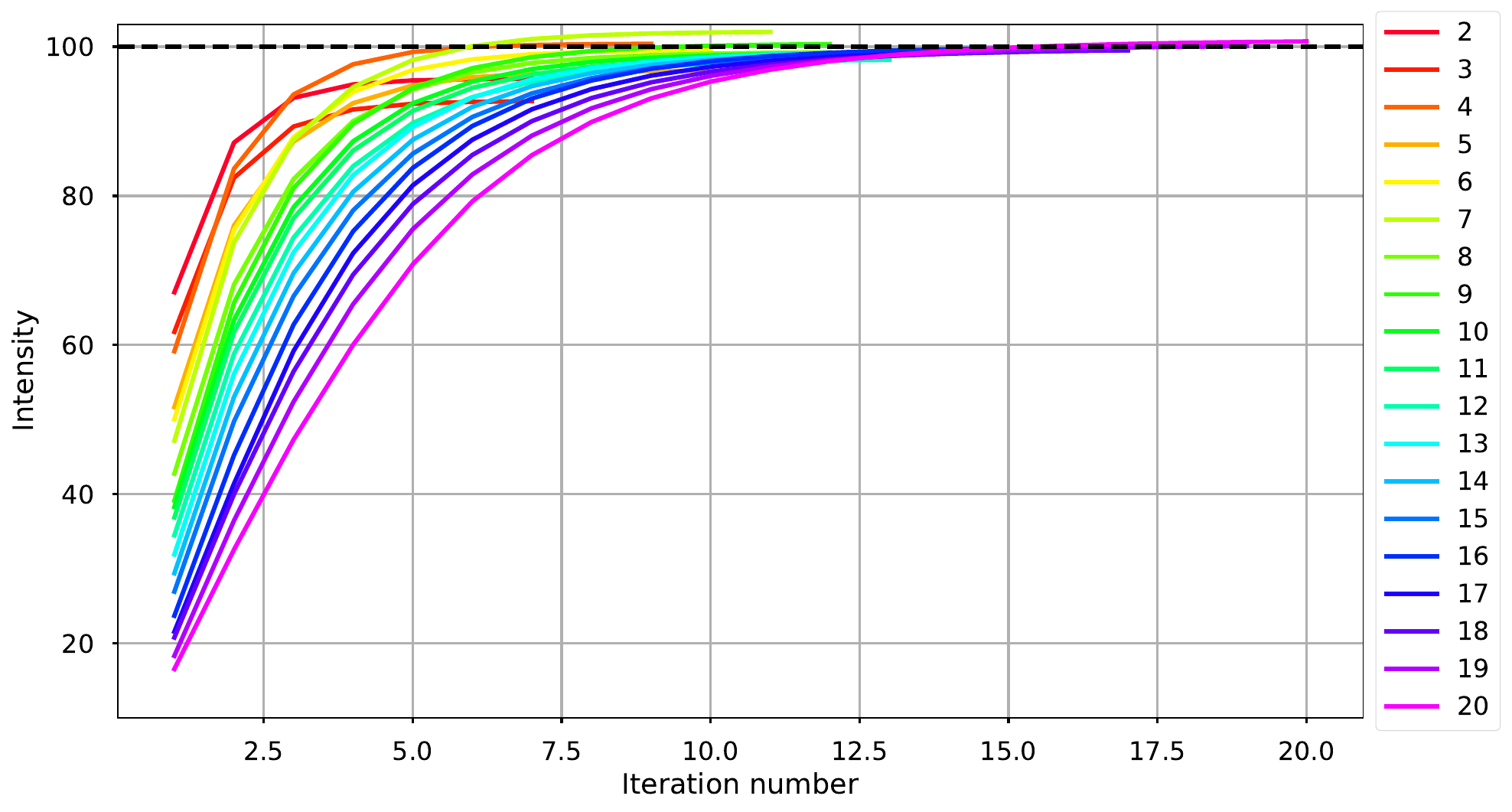}
\caption{AMAT$_\textrm{L2}$ for the flux estimation of an exoplanet. Black dashed line represents the intensity of the injected planet. The algorithm is applied with different ranks ranging from 2 to 20. The iteration number for each rank varies (terminated before reaching the maximum iteration) as the changes in intensity become smaller than the given threshold.}\label{fig:flux_estimation_l2}
\end{figure}

\begin{figure}[!t]
        \centering
        \includegraphics[width=0.48\textwidth]{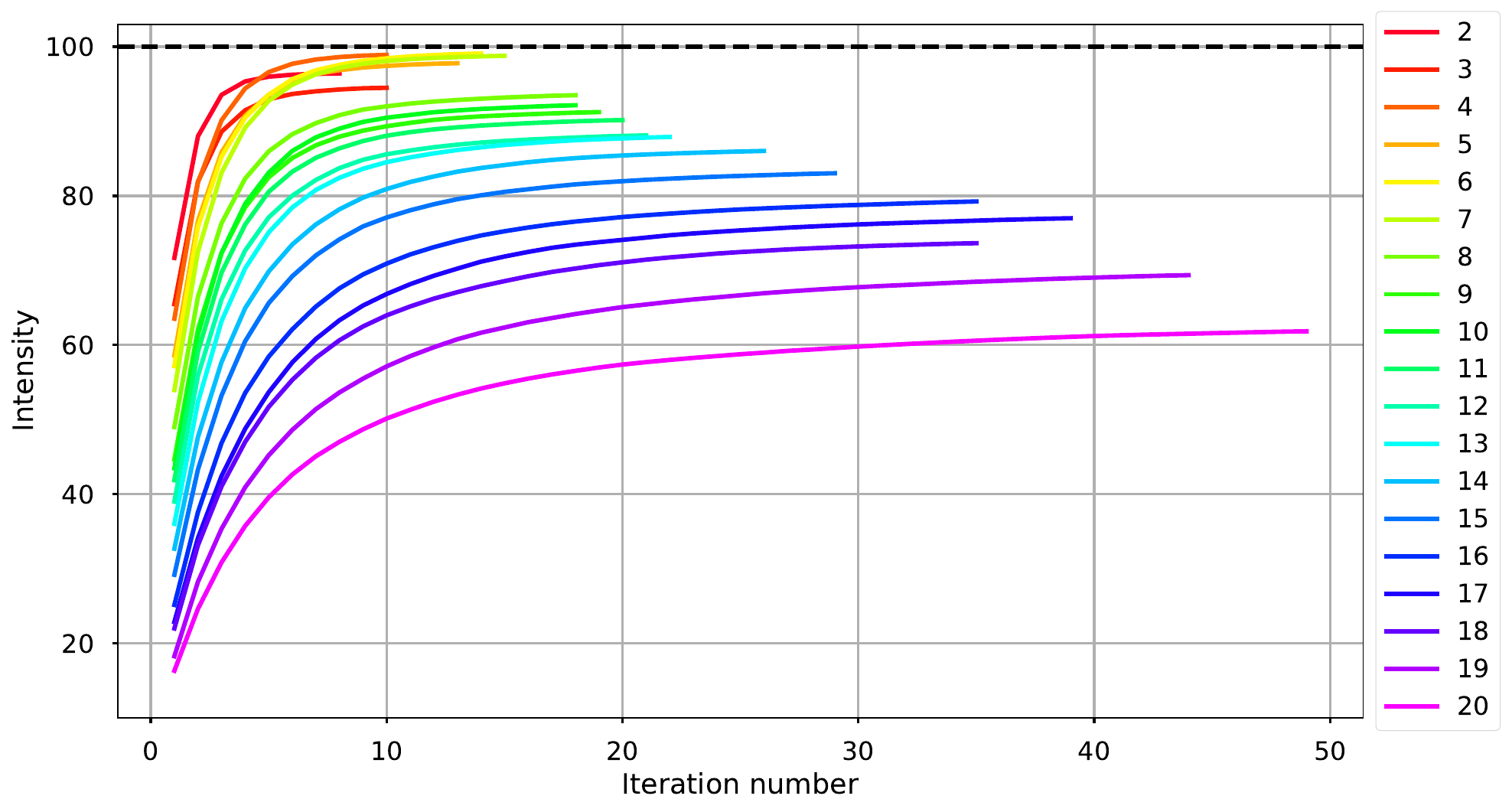}
\caption{AMAT$_\textrm{L1}$ for the flux estimation of an exoplanet. Same inputs as Fig. \ref{fig:flux_estimation_l2}.}\label{fig:flux_estimation_l1}
\end{figure}

\begin{figure*}[!t]
    \centering
    \begin{subfigure}[b]
    {\textwidth}
        \includegraphics[width=\textwidth]{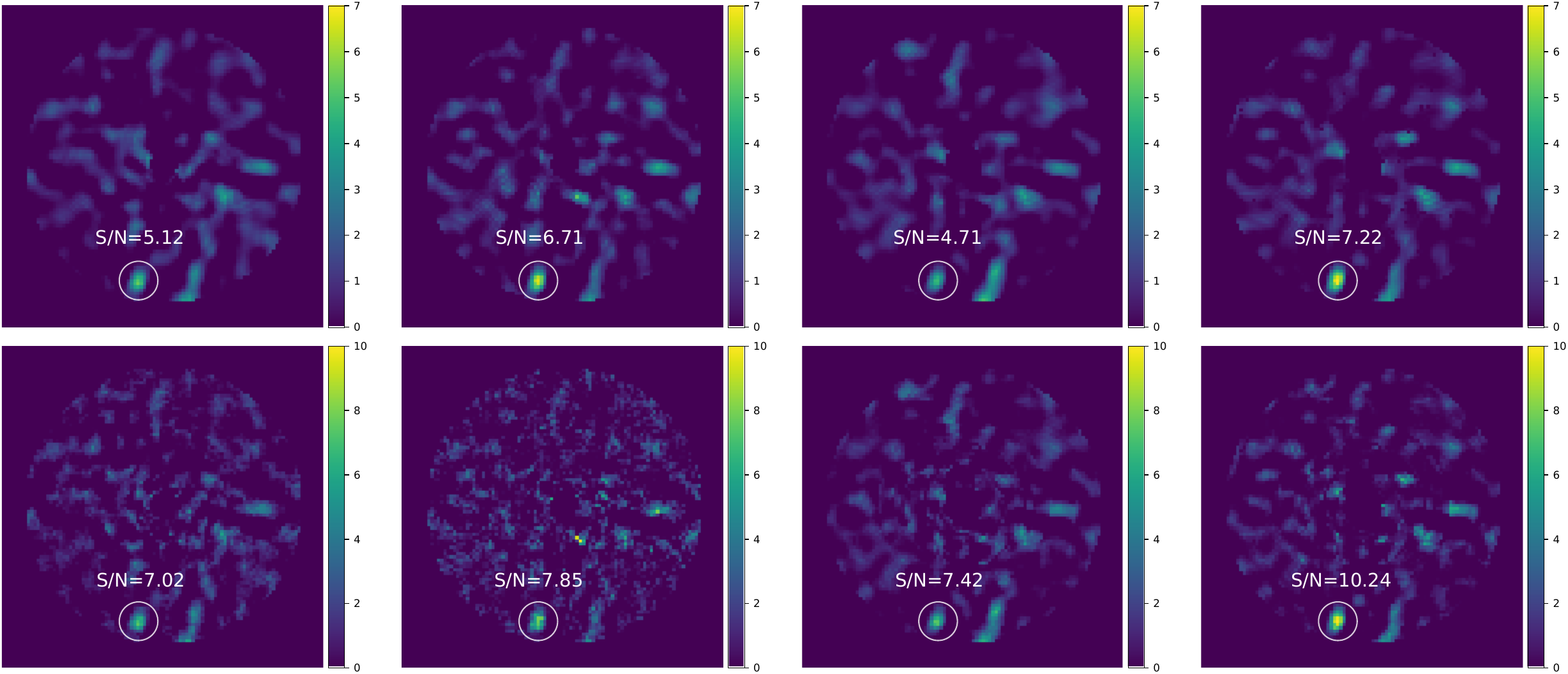}
    \end{subfigure}
    \caption{From left to right: S/N maps after AnnPCA, AnnL1-LRA, AMAT$_\textrm{L2}$, AMAT$_\textrm{L1}$, respectively. Top: S/N maps calculated using VIP package. Bottom: S/N maps proposed by \citet{bonse2023comparing}. The white circles highlight the location of the planet.}\label{fig:snr_lrmap_pca_l1lra_amat}
\end{figure*}

To test this method, we inject a fake planet into the 51 Eri dataset. 
In this scenario, we already know the intensity of the injected planet. In Fig.~\ref{fig:flux_estimation_l2}, we apply the L2 norm version of the AMAT algorithm (AMAT$_\textrm{L2}$) for different ranks from 2 to 20. Although we observe that the algorithm requires more iterations at large ranks, the algorithm approximates the injected intensity for many rank possibilities. Similarly, we apply L1 norm version of the AMAT algorithm (AMAT$_\textrm{L1}$) in Fig.~\ref{fig:flux_estimation_l1}. However, unlike AMAT$_\textrm{L2}$, AMAT$_\textrm{L1}$ only gets close to the injected intensity at small ranks, but not at all when using large ranks. The underlying reason for this is traced back to AMAT$_\text{L1}$'s initialization process; it begins with a randomized SVD for its initial approximation but subsequently relies on the outcomes of preceding steps for further initialization within its iterative process. This dependency introduces a sensitivity to the initial conditions, which coupled with the inability of the algorithm to guarantee recovery of the global optimum, restricts the effectiveness of AMAT$_\text{L1}$ at accurately capturing the injected intensity.

We also aimed to assess the effect of injecting a planet at different positions within the pixel. Initially, we injected a planet with an intensity of 100 at the center of the pixel, which resulted in an intensity of 100.07 after the AMAT$_\textrm{L2}$ algorithm ran. Next, we injected planets with the same intensity at each of the four corners of the pixel in four separate instances. Upon analysis of the trajectories originating from this pixel in the algorithm, the resulting intensities were observed to be 97.23, 96.31, 97.19, and 96.36, respectively. We attribute this flux loss to the intra-pixel variation of flux in the injected PSF, considering the fact that the flux map is only estimated at the center of each pixel.

\section{Performance evaluation}\label{sec:results}

In this section, we perform a comprehensive performance evaluation of the proposed algorithms by using S/N maps for visual comparison, contrast curves to assess sensitivity limits, and receiver operating characteristic (ROC) curves derived from datasets into which synthetic planets were injected. These approaches allow for a detailed comparison of algorithmic effectiveness. We first make use of the SPHERE-IRDIS 51 Eri dataset described in Sect.~\ref{sec:intro}, cropping frames to 100-by-100 pixels to reduce computation time for S/N maps and ROC curves. For contrast curves, we use a larger frame size of 200-by-200 pixels to evaluate performance at angular separations up to the edge of the SPHERE well-corrected field. Our analysis then extends to the Exoplanet Imaging Data Challenge (EIDC) datasets, using S/N maps as a comparison metric. This evaluation strategy is designed to rigorously assess the performance of algorithms for distinguishing planetary signals from noise in different datasets, allowing us to compare them on different well-characterized observational datasets.
Finally, we present a comparative study to showcase the efficacy of the AMAT algorithm in flux estimation in Sect.~\ref{sec:flux_estimation}.

\subsection{S/N and Sensitivity limits}\label{sec:snr_contrastcurve}

To begin our comparison, we aim to assess the performance of various algorithms in detecting the real planet within the 51 Eri dataset. To do so, we compare the S/N map obtained after AnnPCA and after the annular version of the L1-LRA algorithm (AnnL1-LRA) described in Sect.~\ref{sec:l1lra}, with the S/N maps built from the output of the annular version of the 
AMAT$_\textrm{L1}$ and AMAT$_\textrm{L2}$ algorithms. To build these S/N maps, we pave the annulus containing the pixel of interest with non-overlapping apertures, and extract the central pixel of each aperture, as proposed by \citet{bonse2023comparing}. This makes sure that signal and noise are defined in the same way (pixel-wise) for AMAT algorithms, and that the pixels used to build the noise estimation are not correlated. We also obtain S/N maps using the VIP package~\cite{GomezGonzalez2016Lowranka, VIP_HCI}, which is a more common version using the concept of signal-to-noise ratio computation of \cite{mawet2014fundamental}. Figure~\ref{fig:snr_lrmap_pca_l1lra_amat} shows the S/N maps generated using the median frame obtained from AnnPCA and AnnL1-LRA, and the flux map obtained from AMAT algorithms. The S/N map after AMAT$_\textrm{L1}$ outperforms the other three algorithms on both S/N maps versions. For each algorithm, the S/N maps proposed by \citet{bonse2023comparing} yield higher values than the VIP version at the location of the planet. However, in AnnPCA and AnnL1-LRA, this version causes more false positives which can be seen in Appendix \ref{fig:EIDC_snr_apca}-\ref{fig:EIDC_snr_apca_center}. Therefore, we use the VIP package for S/N map after (Ann)PCA and AnnL1-LRA in the following sections of this paper.

\begin{figure}[!t]
    \centering
        \includegraphics[width=0.46\textwidth]{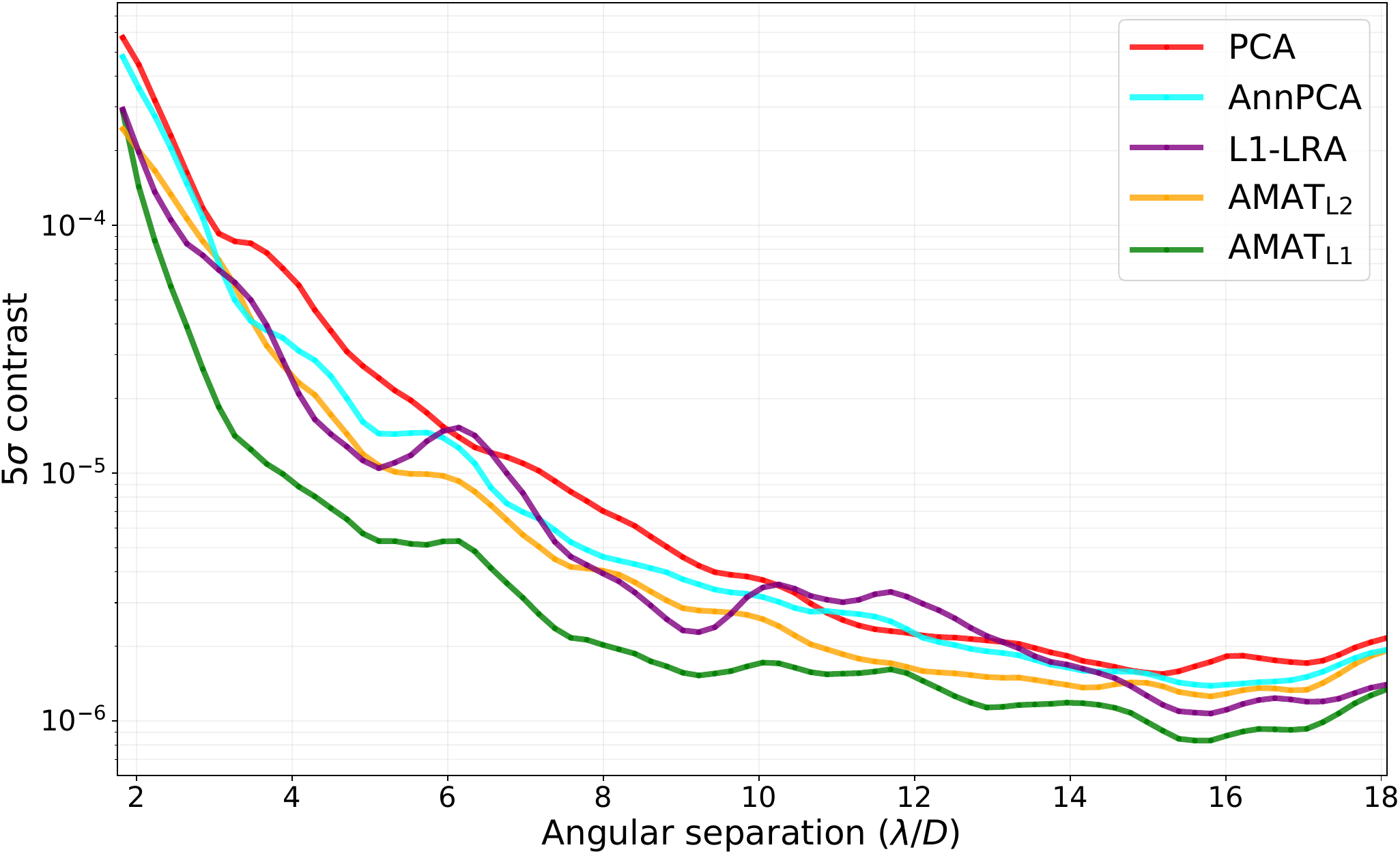}
\caption{Contrast curves for full-frame PCA, AnnPCA and AnnL1-LRA, AMAT$_\textrm{L2}$, and AMAT$_\textrm{L1}$ for the 51 Eri dataset.}\label{fig:contrast_curve}
\end{figure}

A more comprehensive way to assess the sensitivity of various algorithms is to build contrast curves~\citep{mawet2014fundamental}. To do so, we relied on the VIP package. In Fig.~\ref{fig:contrast_curve}, we compare full-frame PCA, AnnPCA, AnnL1-LRA, and our AMAT algorithm using both norms. To generate the AnnPCA contrast curves, we use VIP with its default values except for the rank. We use the same intensities, which are used to obtain the contrast curve of AnnPCA, for the injected planets. Because the different ranks might yield varying results, we applied each algorithm at various ranks, increasing from 5 to 30 in steps of 5. For each algorithm, we selected the deepest contrast curve to represent the best outcome. The optimal ranks were found to be 25 for full-frame PCA, 15 for AnnPCA and AMAT$_\textrm{L2}$, 20 for AnnL1-LRA and AMAT$_\textrm{L1}$. The intensities of the injected planets used to build the contrast curve for each algorithm are based on the noise values in each annulus obtained when applying AnnPCA with default values other than rank. This ensured that the contrast curves were obtained using the dataset with the same injected planets.
As can be observed in Fig.~\ref{fig:contrast_curve}, the performance using full-frame PCA tends to be the worst among the compared methods. It is followed by AnnPCA, which shows slightly better performance. AMAT$_\textrm{L2}$ outperforms AnnPCA by a small margin. The performance of AnnL1-LRA fluctuates based on the separation, indicating a performance where it sometimes outperforms or underperforms compared to the other methods. Finally, the AMAT$_\textrm{L1}$ algorithm demonstrates increased efficacy, outperforming other algorithms across all separations.

\subsection{ROC curves}\label{sec:roc_curves_snr}

While S/N and contrast curves are useful to illustrate the gain provided by AMAT, they do not explore the behaviour of the algorithms as a function of the detection threshold, which can be done through ROC curves. Building ROC curves relies on injecting a large quantity of synthetic planets into the chosen data set. Our process begins with the removal of the real planet present in the dataset using the VIP package \citep{GomezGonzalez2017VIP, VIP_HCI}. 
Subsequently, we inject synthetic planets with an intensity of 1.5 times the standard deviation of the values in the cube at a distance of 2$\lambda/D$ from the star. The injections are placed methodically, starting from 0 to 360 degrees in increments of 3.6 degrees and placing a synthetic planet per scenario. This approach results in a total of 100 different cases for evaluation, effectively covering the entire 360-degree span around the star. 
For each scenario, we apply the algorithms and then examine the location where the planet was injected. A detection within a specified aperture exceeding a predefined threshold is counted as a true positive (TP); then, we check the other apertures for the presence of the signal. If a signal above the threshold is found within these apertures, it is classified as a false positive (FP); the absence of such a signal results in a true negative (TN) classification. This examination across all apertures facilitates the construction of a ROC curve.
Given the critical importance of maintaining a low number of FPs in exoplanet detection, our analysis primarily focuses on achieving a high true positive rate (TPR) without incurring false positives. To better visualize and compare ROC curves, especially to highlight the performance at minimal false positive rate (FPR), we employ a transformed plot of the square root of TPR versus FPR, allowing for an enhanced representation of algorithmic efficiency.

\begin{figure}[!t]
    \centering
        \includegraphics[width=0.48\textwidth]{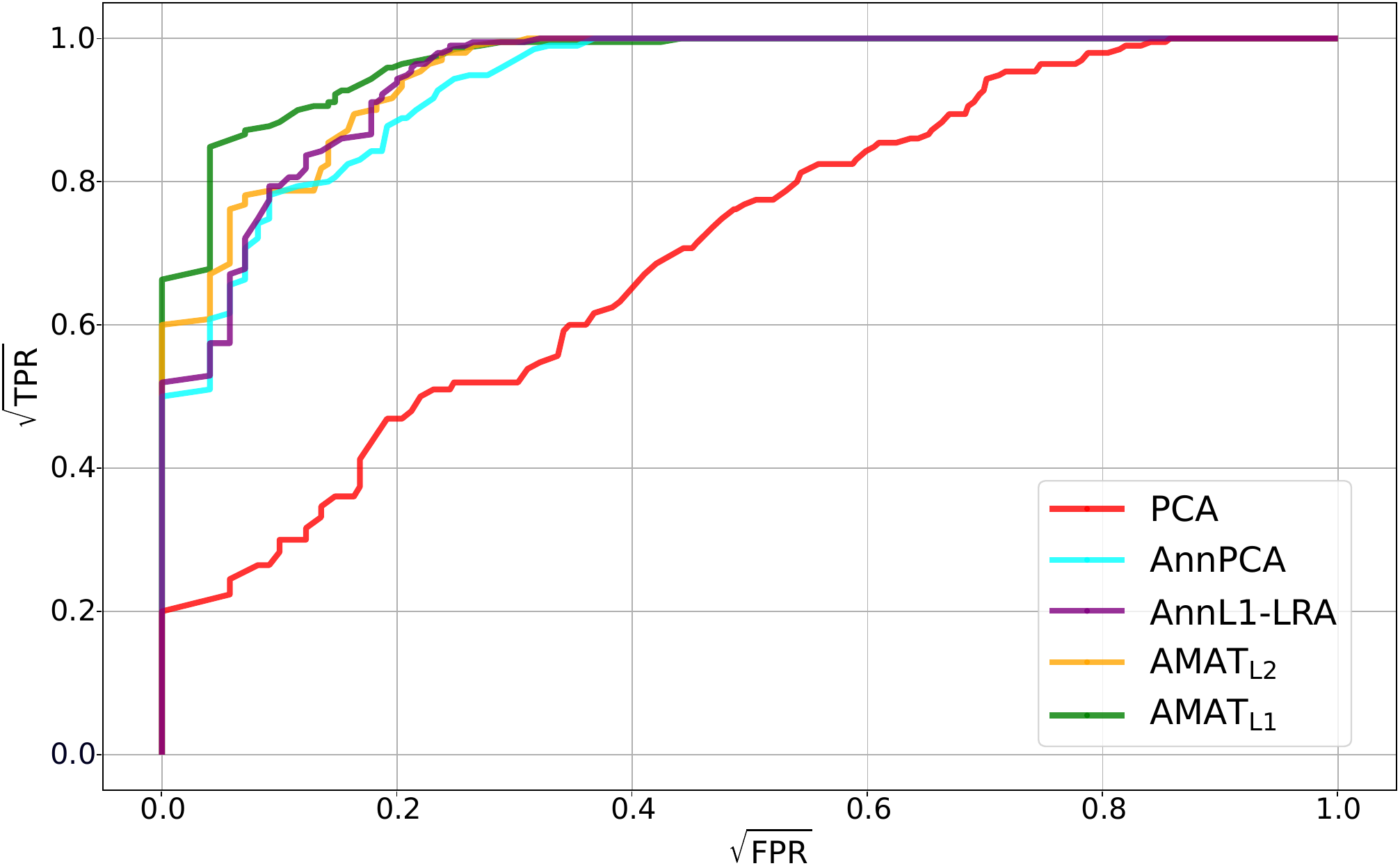}
\caption{ROC curve of S/N maps. We compare full-frame PCA, AnnPCA and AnnL1-LRA, AMAT$_\textrm{L2}$, and AMAT$_\textrm{L1}$ for the 51 Eri dataset. We use the square root to scale the axes in order to better see the low FPR regime. }\label{fig:roc_curve}
\end{figure}

We compare our AMAT algorithms with the results of the full-frame PCA, AnnPCA, and AnnL1-LRA algorithms. 
We use the ranks where we get the best contrast curves for each algorithm. The results displayed in Fig.~\ref{fig:roc_curve} show that our method consistently outperforms the results of full-frame PCA, AnnPCA, and AnnL1-LRA in terms of ROC curves. Moreover, similar to the findings for the S/N map, the results obtained using the L1 norm are better than those obtained with the L2 norm. Furthermore, AnnL1-LRA shows slightly better performance compared to both AnnPCA and full-frame PCA.

\subsection{EIDC results}
As part of our evaluation process, we utilize the datasets from the Exoplanet Imaging Data Challenge \citep[EIDC,][]{cantalloube2020exoplanet}, as specified in Table~\ref{table:datasets} to provide varying datasets for comparing and assessing our algorithm alongside state-of-the-art HCI algorithms. This challenge encompasses a diverse array of ADI sequences, including nine ADI datasets, with a total of 20 injected planet signals. These signals exhibit varying contrasts and positional coordinates to create an evaluation from three different instruments: VLT/SPHERE-IRDIS, Keck/NIRC2, and LBT/LMIRCam each providing three datasets.

\begin{figure*}[!t]
    \centering
    \begin{subfigure}[b]
    {0.32\textwidth}
        \includegraphics[width=\textwidth]{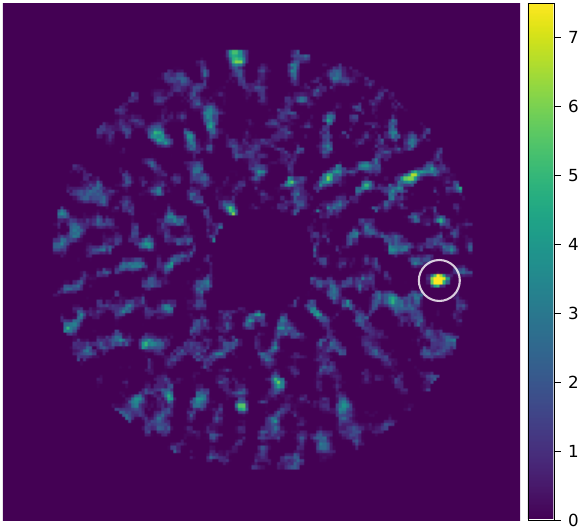}
        \caption{sph1}
    \end{subfigure}
    \begin{subfigure}[b]
    {0.32\textwidth}
        \includegraphics[width=\textwidth]{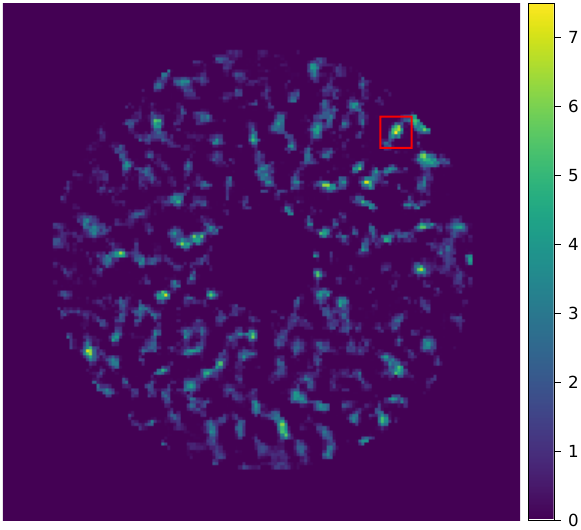}
        \caption{sph2}
    \end{subfigure}
    \begin{subfigure}[b]
    {0.32\textwidth}
        \includegraphics[width=\textwidth]{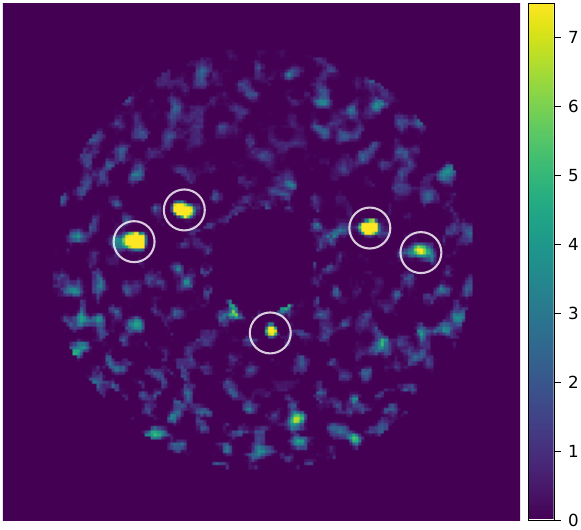}
        \caption{sph3}
    \end{subfigure}
    \begin{subfigure}[b]
    {0.32\textwidth}
        \includegraphics[width=\textwidth]{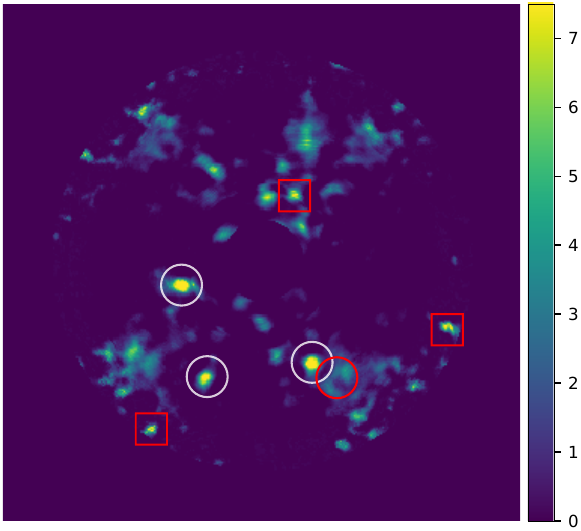}
        \caption{nrc1}
    \end{subfigure}
    \begin{subfigure}[b]
    {0.32\textwidth}
        \includegraphics[width=\textwidth]{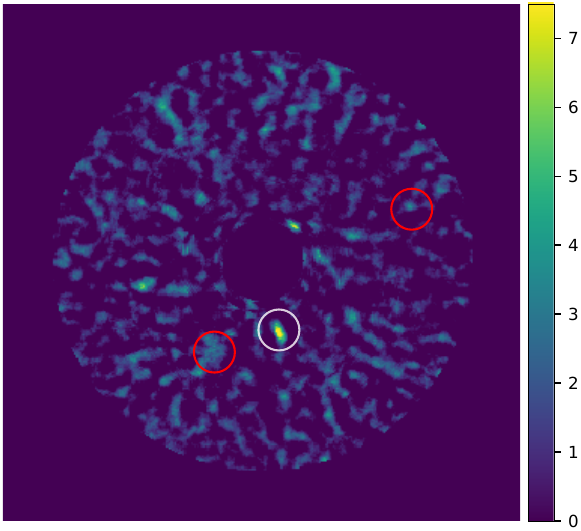}
        \caption{nrc2}
    \end{subfigure}
    \begin{subfigure}[b]
    {0.32\textwidth}
        \includegraphics[width=\textwidth]{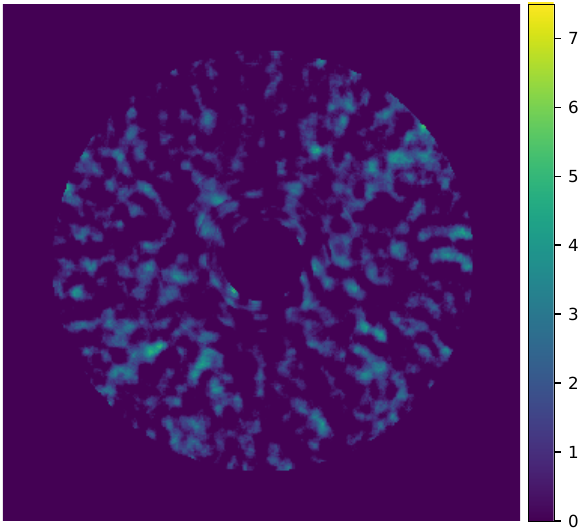}
        \caption{nrc3}
    \end{subfigure}
    \begin{subfigure}[b]
    {0.32\textwidth}
        \includegraphics[width=\textwidth]{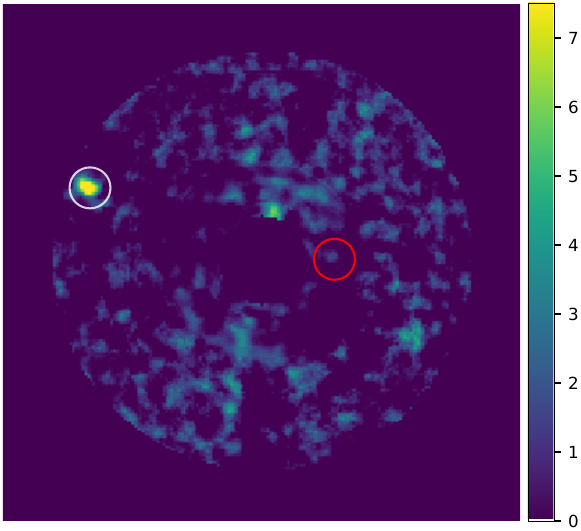}
        \caption{lmr1}
    \end{subfigure}
    \begin{subfigure}[b]
    {0.32\textwidth}
        \includegraphics[width=\textwidth]{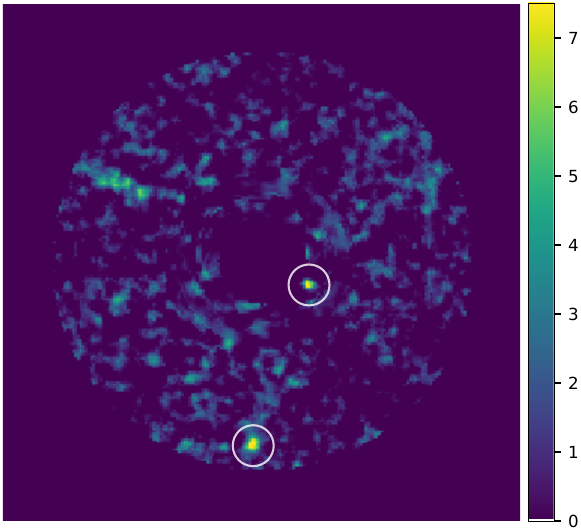}
        \caption{lmr2}
    \end{subfigure}
    \begin{subfigure}[b]
    {0.32\textwidth}
        \includegraphics[width=\textwidth]{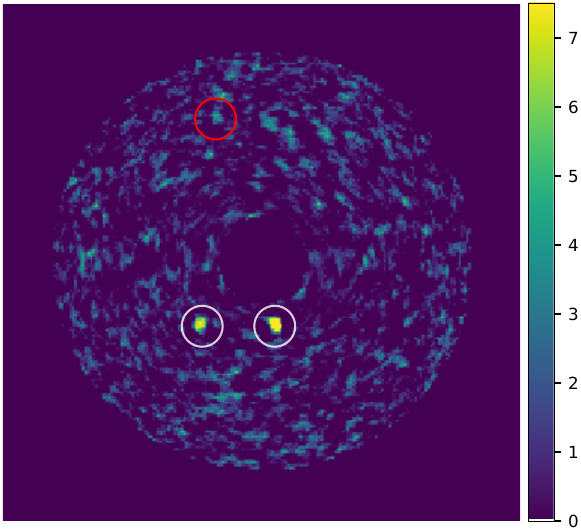}
        \caption{lmr3}
    \end{subfigure}
\caption{S/N maps after AMAT$_\textrm{L1}$: In these S/N maps, white circles represent TP, red squares denote FP, and red circles signify FN.}\label{fig:EIDC_snr_amatl1}
\end{figure*}

\begin{figure*}[!t]
    \centering
    \begin{subfigure}[b]
    {0.32\textwidth}
        \includegraphics[width=\textwidth]{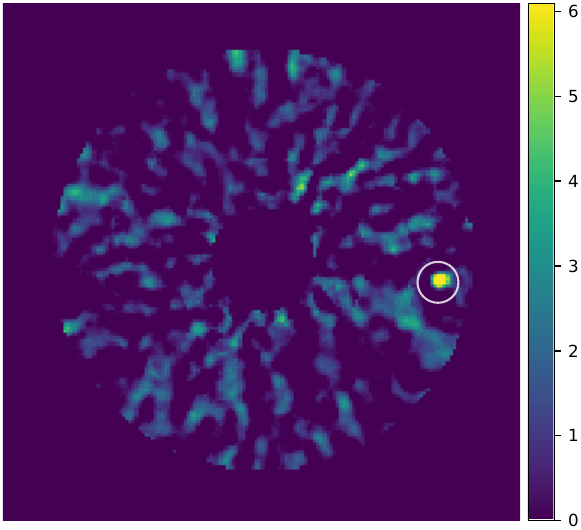}
        \caption{sph1}
    \end{subfigure}
    \begin{subfigure}[b]
    {0.32\textwidth}
        \includegraphics[width=\textwidth]{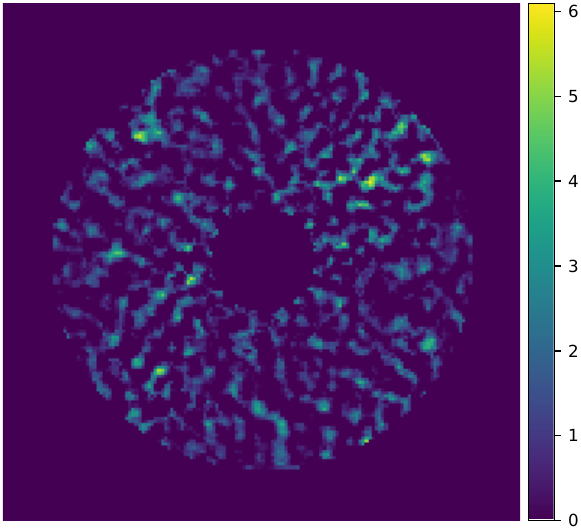}
        \caption{sph2}
    \end{subfigure}
    \begin{subfigure}[b]
    {0.32\textwidth}
        \includegraphics[width=\textwidth]{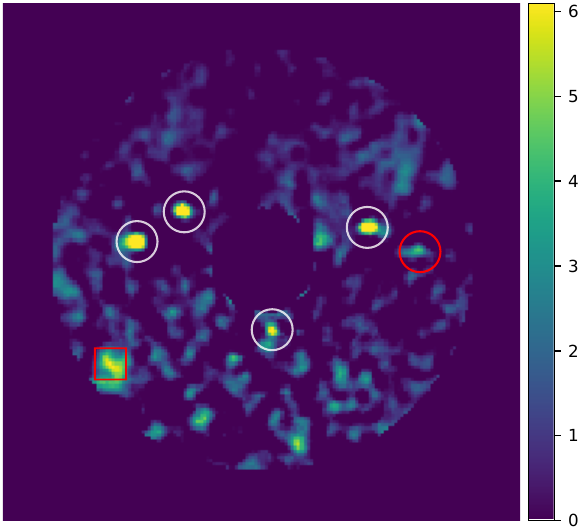}
        \caption{sph3}
    \end{subfigure}
    \begin{subfigure}[b]
    {0.32\textwidth}
        \includegraphics[width=\textwidth]{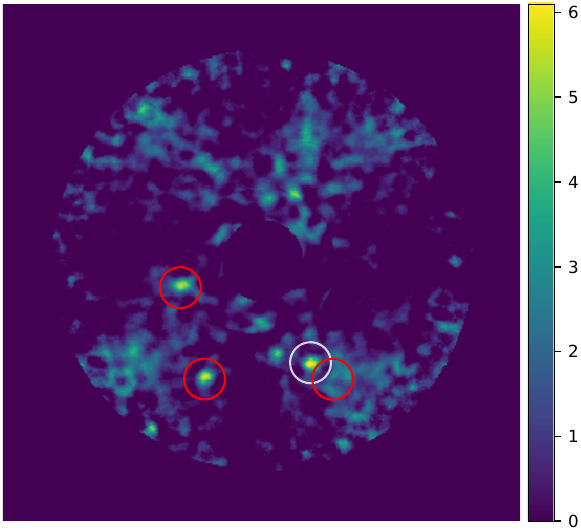}
        \caption{nrc1}
    \end{subfigure}
    \begin{subfigure}[b]
    {0.32\textwidth}
        \includegraphics[width=\textwidth]{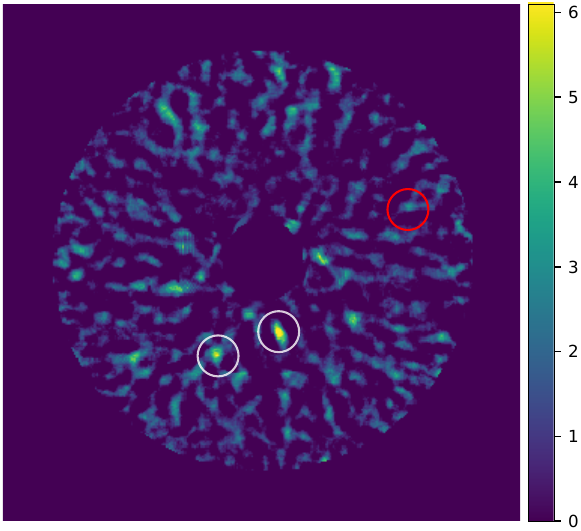}
        \caption{nrc2}
    \end{subfigure}
    \begin{subfigure}[b]
    {0.32\textwidth}
        \includegraphics[width=\textwidth]{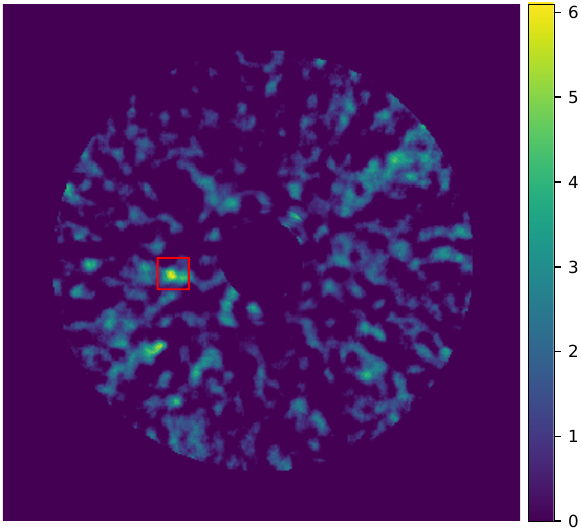}
        \caption{nrc3}
    \end{subfigure}
    \begin{subfigure}[b]
    {0.32\textwidth}
        \includegraphics[width=\textwidth]{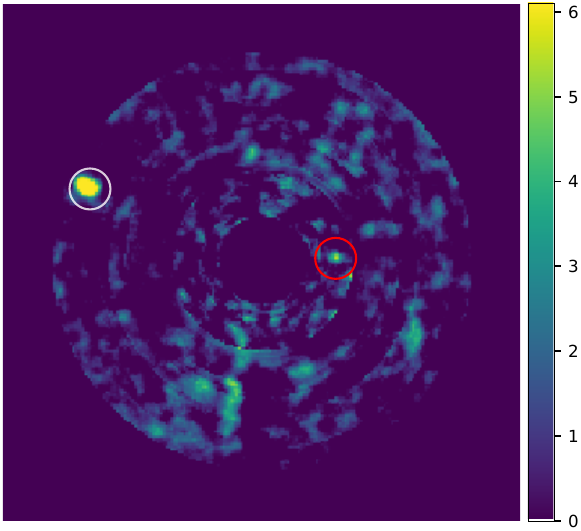}
        \caption{lmr1}
    \end{subfigure}
    \begin{subfigure}[b]
    {0.32\textwidth}
        \includegraphics[width=\textwidth]{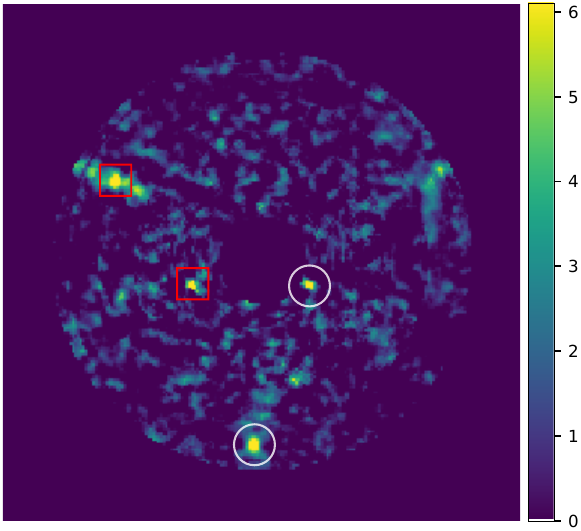}
        \caption{lmr2}
    \end{subfigure}
    \begin{subfigure}[b]
    {0.32\textwidth}
        \includegraphics[width=\textwidth]{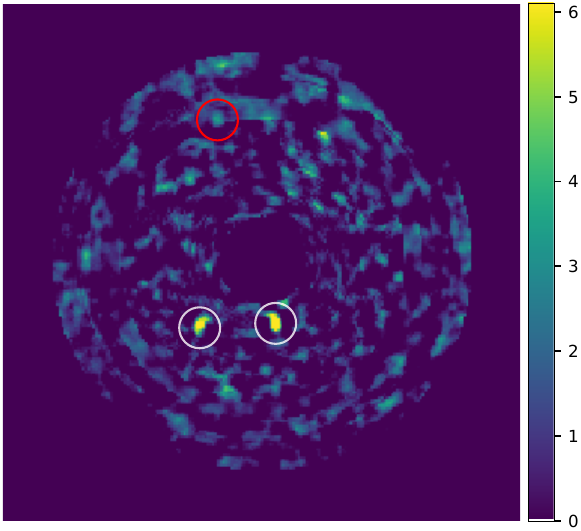}
        \caption{lmr3}
    \end{subfigure}
\caption{S/N maps after AMAT$_\textrm{L2}$: In these S/N maps, white circles represent TP, red squares denote FP, and red circles signify FN.}\label{fig:EIDC_snr_amatl2}
\end{figure*}

As with the EIDC, our assessment involves the generation of detection maps by each algorithm for every ADI sequence. A detection map refers to applying a threshold for detection to the S/N maps or likelihood ratio maps detailed in Sect.~\ref{sec:likelihood}. In the result report of EIDC, a set of standard metrics was employed to compare the performance of these detection maps. 
Therefore, to apply a common metric for comparing the EIDC algorithms, we calculate F1-scores using the same definition
\begin{align}
    \textrm{F1-score} = \frac{2\textrm{TP}}{2\textrm{TP}+\textrm{FP}+\textrm{FN}}.
\end{align}

To determine the values of TP, FP, and FN, we need to decide on a threshold value, as this can significantly impact performance. To determine the appropriate threshold, we inject synthetic planets into the 51 Eri dataset. We create three distinct datasets illustrated in Appendix~\ref{sec:eri_fake}, each injected with three, two, and four planets respectively, placed at different locations, and subsequently run our algorithm. We experiment with varying thresholds to identify the threshold yielding the highest F1-score. Fig.~\ref{fig:f1_threshold_snr} illustrates the plot of F1-score against threshold for the 51 Eri dataset with synthetic planets. Based on this analysis, if we select a threshold between 6.6-7.7 for S/N maps after AMAT$_\textrm{L1}$, we can detect all planets without false positives and 
 this allows us to achieve the highest F1-score.

In the published results report, we have knowledge of the locations of the planets and evaluated various threshold values as evidenced by the F1 scores presented in Fig.~\ref{fig:f1_threshold}. Since we also established our threshold range by testing injections into the 51 Eri dataset, we conducted a posteriori verification to ensure that the thresholds also yield a high F1 score when applied to the EIDC datasets.  Participating in the EIDC, where we had multiple opportunities to submit and observe the F1-score, allows us to undergo a similar process. Selecting the value that yields the highest F1-score represents a fair and comparable approach. In our comparison, we used the threshold of 7.5 within the 6.6-7.7 range which we obtain the highest F1-score for the 51 Eri dataset.

The effectiveness of the AMAT$_\textrm{L1}$ algorithm is illustrated in Fig.~\ref{fig:EIDC_snr_amatl1} through S/N maps.  The analysis of the VLT/SPHERE-IRDIS datasets showcases the proficiency of the algorithm, where it successfully identified all six planets, with only one false positive. In the Keck/NIRC2 dataset, the algorithm accurately detected four out of seven planets with three false positives, and in the LBT/LMIRCam dataset five out of seven planets without any false positives. 
The results of our experiments using AnnPCA on the EIDC data set are shown in Fig.~\ref{fig:EIDC_snr_apca} for reference, to illustrate the significant gain provided by AMAT in terms of TPR. Additionally, the AMAT$_\textrm{L2}$ algorithm exhibits a performance between AMAT$_\textrm{L1}$ and AnnPCA in Fig.~\ref{fig:EIDC_snr_amatl2}. It detects five out of six planets with one false positive in the VLT/SPHERE-IRDIS datasets, three out of seven planets in the Keck/NIRC2 dataset with one false positive, and five out of seven planets with two false positives in the LBT/LMIRCam dataset. We compute F1-scores based on these maps and compare them with the best results of the published results report in Fig.~\ref{fig:f1_threshold}.

\subsection{Flux estimation performance}\label{sec:flux_estimation}
\begin{figure*}[!t]
    \centering
    \includegraphics[width=\textwidth]{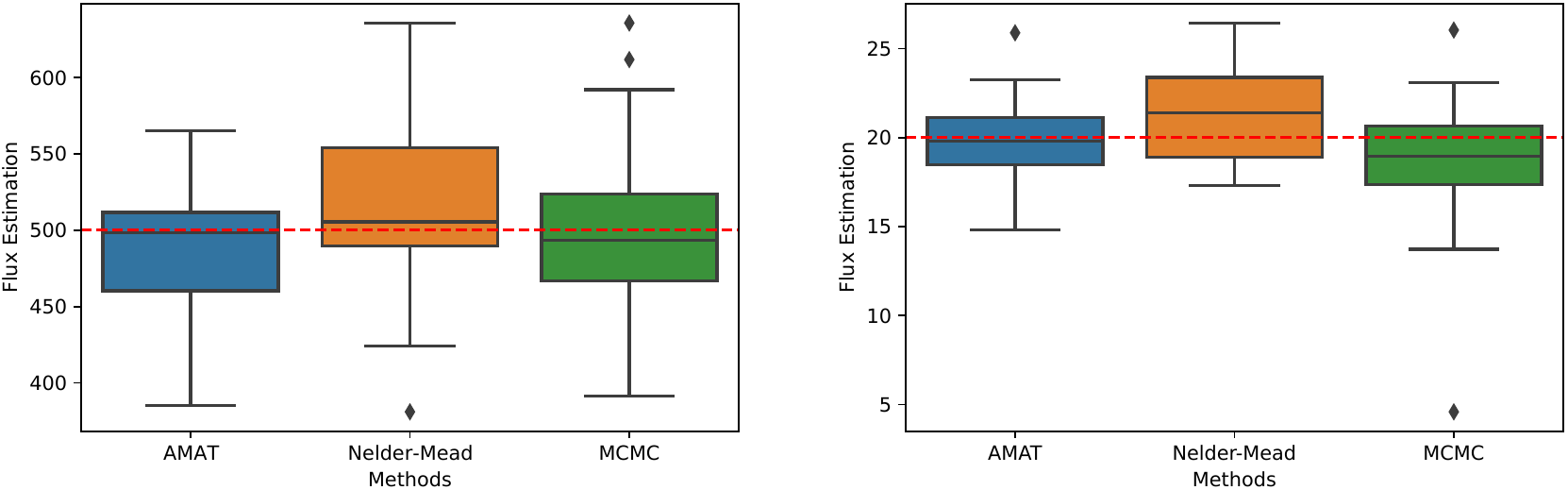}
\caption{Box plot for flux estimation for planets which are injected in 4$\lambda/D$ (left) and 10$\lambda/D$ (right) separation. Dashed red line represents the injected flux value.}\label{fig:box_plot_small}
\end{figure*}
For the purpose of flux estimation, we conducted a comparative analysis of our AMAT$_\textrm{L2}$ algorithm against the NEGFC method, which uses AnnPCA, implemented in VIP, where the companion parameters are estimated either through a Nelder Mead minimization method or through a Markov Chain Monte Carlo (MCMC) method~\citep{wertz2017vlt}. Some other forward modeling or inverse problem approaches \citep{cantalloube2015direct, wang2016orbit} also have the potential to provide accurate flux measurements and may outperform both NegFC and AMAT. However, a full comparison of these methods is beyond the scope of this paper. We designed two distinct sets of comparisons, one with the synthetic planets injected at a small separation (4$\lambda/D$) and the other at a large separation (10$\lambda/D$).
In both scenarios, we created various cases by rotating the position of the injected planet from 0 to 350 degrees in increments of 10 degrees. This resulted in 36 different cases for each scenario. This systematic approach was applied to both the 4$\lambda/D$ and the 10$\lambda/D$ separations, thereby ensuring a comprehensive evaluation. Planets were injected at specified radii and angles without consideration for whether the location falls precisely on the center or the edge of the pixel, ensuring consistency across all cases.
To decide on the rank for each algorithm, we use the approach suggested in VIP~\citep{VIP_HCI,GomezGonzalez2017VIP}, which finds the optimal number of principal components in terms of S/N using PCA. In the AMAT algorithm, after a maximum of 50 iterations, or when the relative change of the intensity $a_g^{(i)}$ is minor, \emph{i.e.}, when $|a_g^{(i)} - a_g^{(i-1)}|/|a_g^{(i)}| < 10^{-3}$, we simply stop the algorithm.

In Fig.~\ref{fig:box_plot_small}, the AMAT$_\textrm{L2}$ algorithm demonstrates excellent accuracy in flux estimations for both small and large separations, as evidenced by the median line of the boxplots precisely aligning with the injected intensities. Contrastingly, the boxplots for the Nelder-Mead and MCMC methods exhibit the presence of outliers, indicating lower precision in their flux estimations. Specifically, the Nelder-Mead method tends to produce higher flux values, often deviating significantly from the injected values, while the MCMC method frequently results in lower flux estimations.

\section{Improving detection performance with likelihood ratio maps}\label{sec:likelihood}

In \citet{daglayan2022likelihood}, we presented an enhanced approach for exoplanet detection utilizing likelihood ratio maps (LRM), which offers improvements over traditional S/N maps. The LRM is derived through a maximum likelihood estimation, employing Laplacian distributions. It consists of the ratio of the maximum likelihood $\mathcal{L}_g(\hat{a}_g | R)$ to the likelihood of the null hypothesis $\mathcal{L}_g(0 | R)$, which corresponds to the absence of a planet
\begin{equation}
        \begin{split}
	\log \Lambda_g(R) &= \log \left(\frac{
		 \mathcal{L}_g(\hat{a}_g | R)
	}{
	 \mathcal{L}_g(0 | R)
	}\right) \\
	&= -\sum_{(\theta, r) \in {\Omega}_g}\!\!
		\frac{
			|R(\theta, r) - \hat{a}_g P_g(\theta, r)|
			- |{R(\theta, r)}|
		}{
			\sigma_{R}(r)
		}
       \end{split}
\end{equation}
where $R$ is the residual cube and $\hat{a}_g$ is the estimated flux by solving the following optimization problem
\begin{align} 
\label{eq:mc_lap_loglike_optim}
    \begin{split}
	\hat{a}_g
	&= \mathrm{argmax}_a \log \mathcal{L}_g(a|R) \\
	&= \mathrm{argmin}_a \sum_{(\theta, r) \in {\Omega}_g}\!\!\!\!
	\frac{|{R(\theta, r) - a P_g(\theta, r)}|}{\sigma_{R}(r)}.
    \end{split}
\end{align}
In light of this information, we have combined the AMAT algorithm with the LRM. After obtaining the residual cube, we apply the LRM algorithm to our results. 

\begin{figure}[!t]
    \centering
        \includegraphics[width=0.48\textwidth]{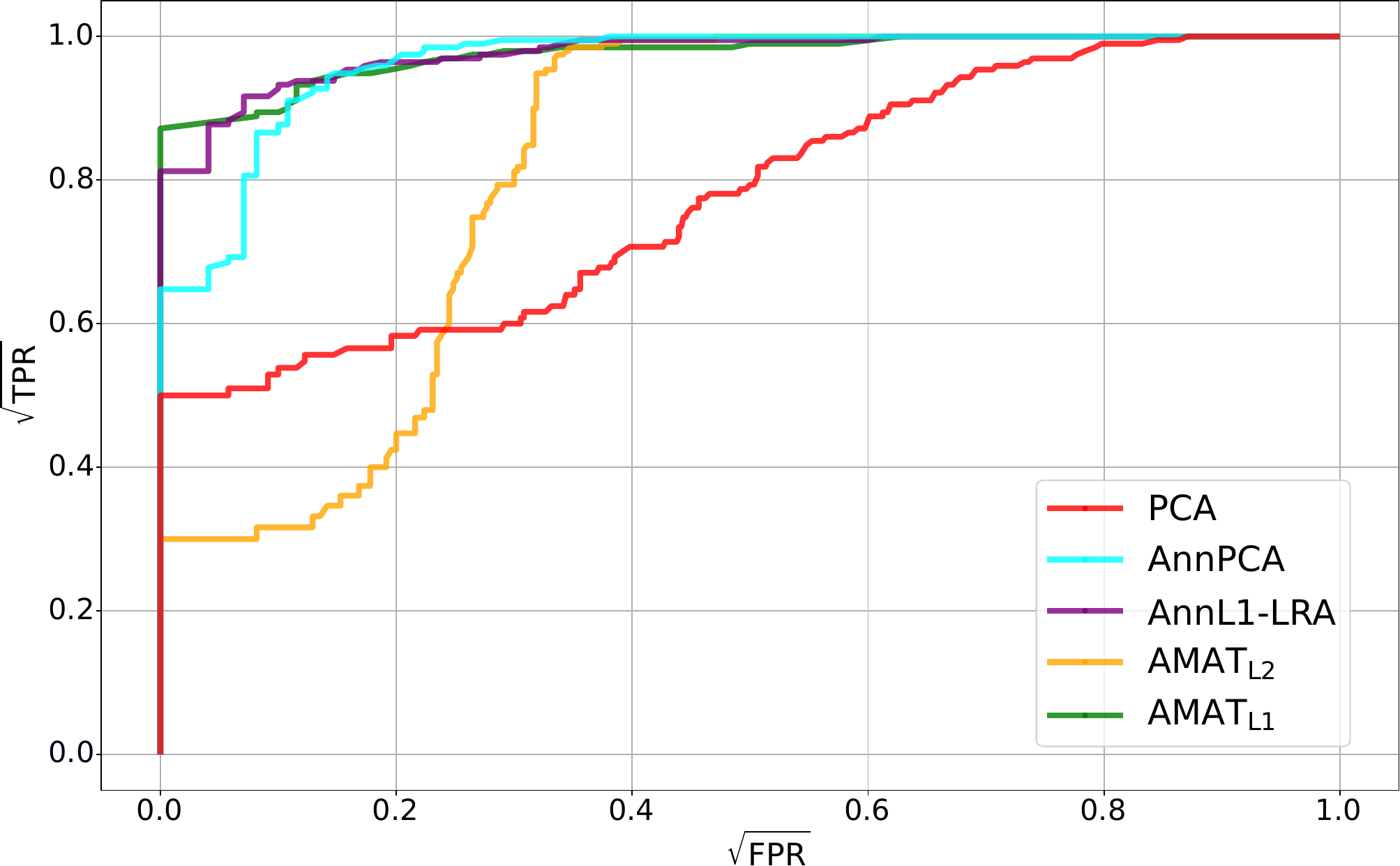}
\caption{ROC curve of LRMs. We compare full-frame PCA, AnnPCA and AnnL1-LRA, AMAT$_\textrm{L2}$, and AMAT$_\textrm{L1}$. }\label{fig:roc_curve_lr}
\end{figure}

We obtained the ROC curve using the residual cubes presented in Sect.~\ref{sec:roc_curves_snr} and applied the LRM instead of the S/N map in Fig. \ref{fig:roc_curve_lr}. Each algorithm, except AMAT$_\textrm{L2}$, demonstrated higher performance compared to the S/N map results in Fig.~\ref{fig:roc_curve}. We believe this discrepancy arises from AMAT$_\textrm{L2}$ being based on a Gaussian distribution, while LRM relies on a Laplacian distribution. Among the algorithms, AMAT$_\textrm{L1}$ showed the best performance, while AnnL1-LRA also exhibited strong results. Both algorithms are Laplacian-based, which likely contributes to their superior performance.

To decide on the threshold, we used the same method as described in Sect.~\ref{sec:roc_curves_snr} for finding the threshold of S/N maps. Based on this analysis, if we choose a value between 67-77 for the LRMs after AMAT$_\textrm{L1}$, we get the highest F1 score for the 51 Eri dataset, and as for the case of AMAT-S/N, we check that the threshold for LRMs after AMAT$_\textrm{L1}$ is indeed in this range to maximize the F1 score on the EIDC.

\begin{figure*}[!t]
    \centering
    \begin{subfigure}[b]
    {0.32\textwidth}
        \includegraphics[width=\textwidth]{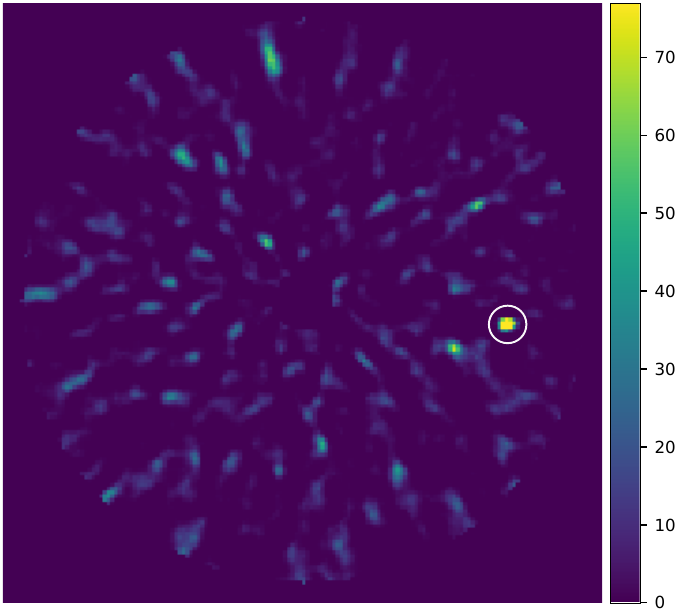}
        \caption{sph1}
    \end{subfigure}
    \begin{subfigure}[b]
    {0.32\textwidth}
        \includegraphics[width=\textwidth]{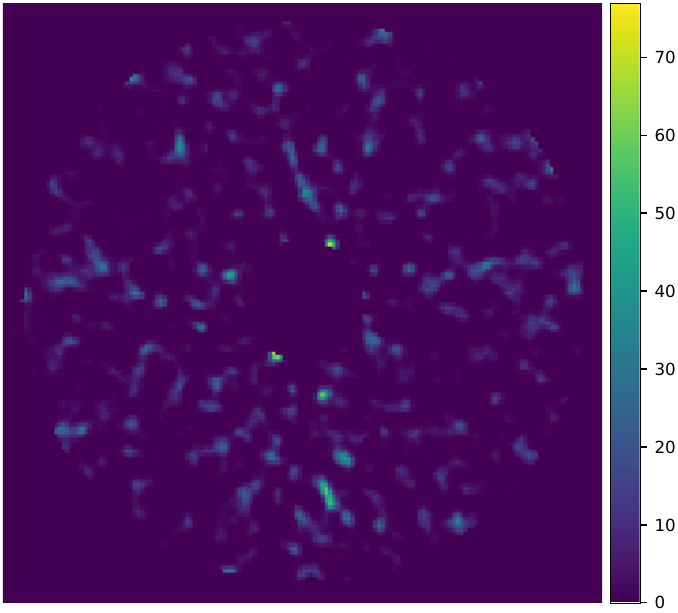}
        \caption{sph2}
    \end{subfigure}
    \begin{subfigure}[b]
    {0.32\textwidth}
        \includegraphics[width=\textwidth]{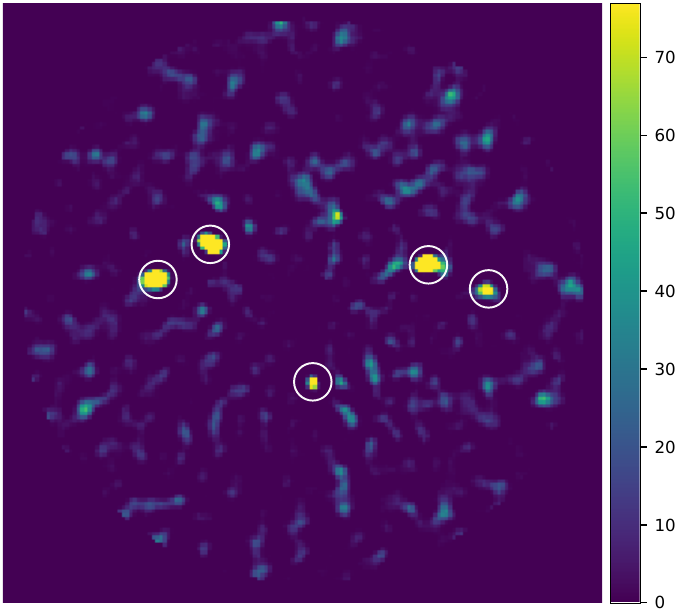}
        \caption{sph3}
    \end{subfigure}
    \begin{subfigure}[b]
    {0.32\textwidth}
        \includegraphics[width=\textwidth]{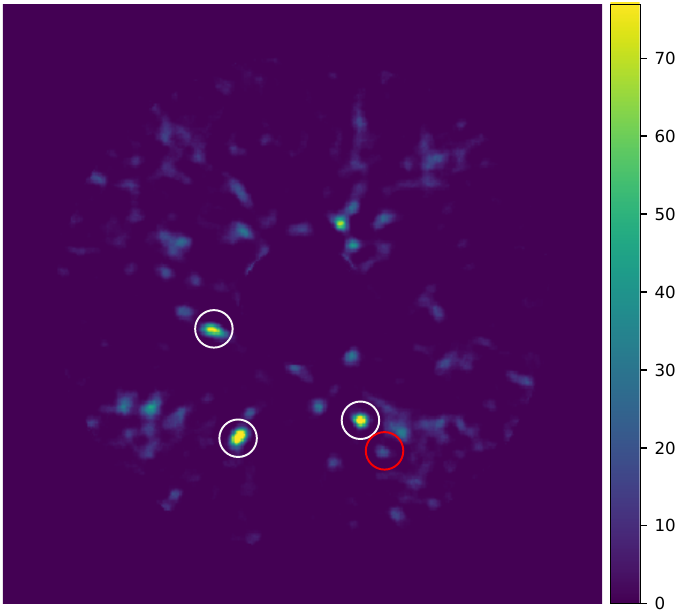}
        \caption{nirc1}
    \end{subfigure}
    \begin{subfigure}[b]
    {0.32\textwidth}
        \includegraphics[width=\textwidth]{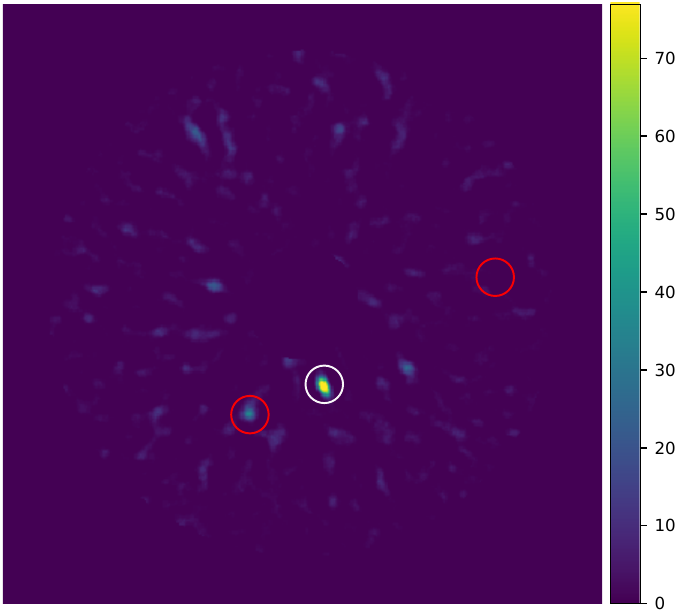}
        \caption{nirc2}
    \end{subfigure}
    \begin{subfigure}[b]
    {0.32\textwidth}
        \includegraphics[width=\textwidth]{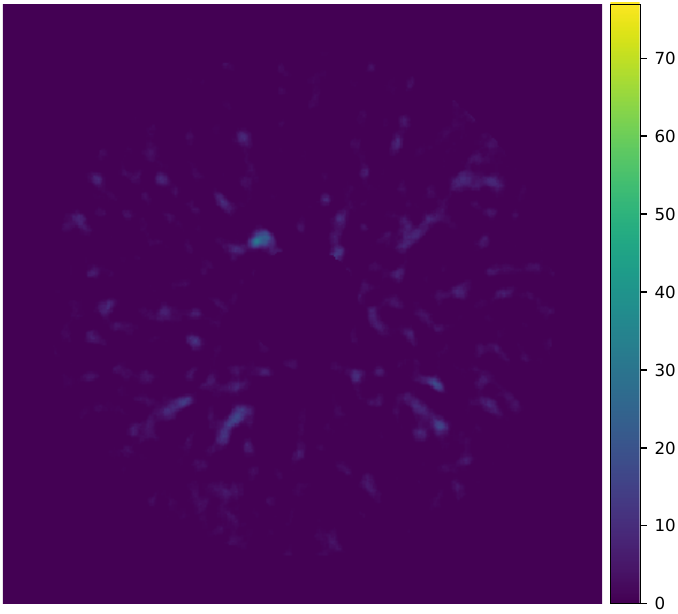}
        \caption{nirc3}
    \end{subfigure}
    \begin{subfigure}[b]
    {0.32\textwidth}
        \includegraphics[width=\textwidth]{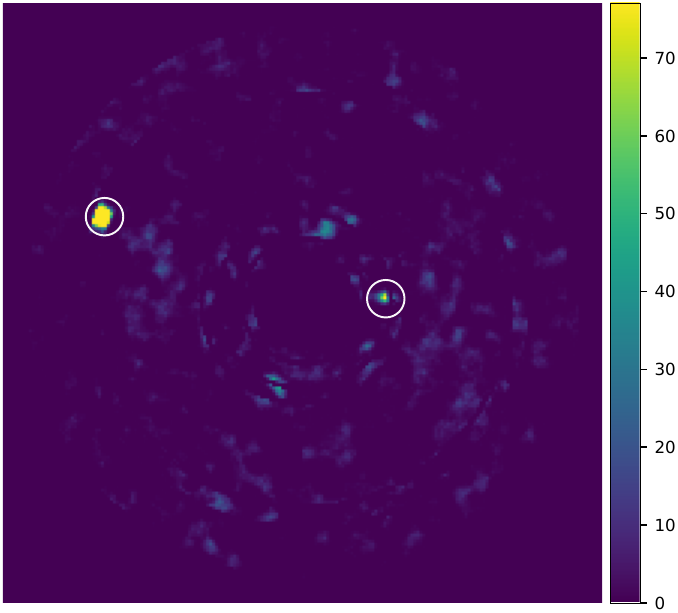}
        \caption{lmr1}
    \end{subfigure}
    \begin{subfigure}[b]
    {0.32\textwidth}
        \includegraphics[width=\textwidth]{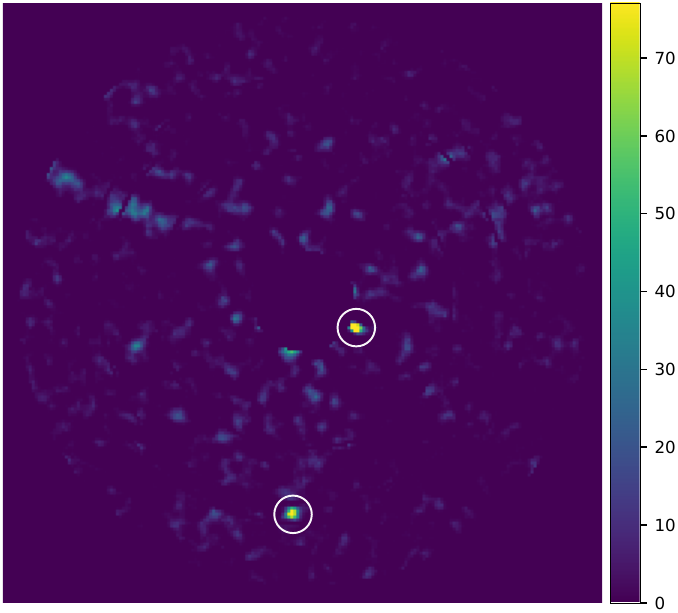}
        \caption{lmr2}
    \end{subfigure}
    \begin{subfigure}[b]
    {0.32\textwidth}
        \includegraphics[width=\textwidth]{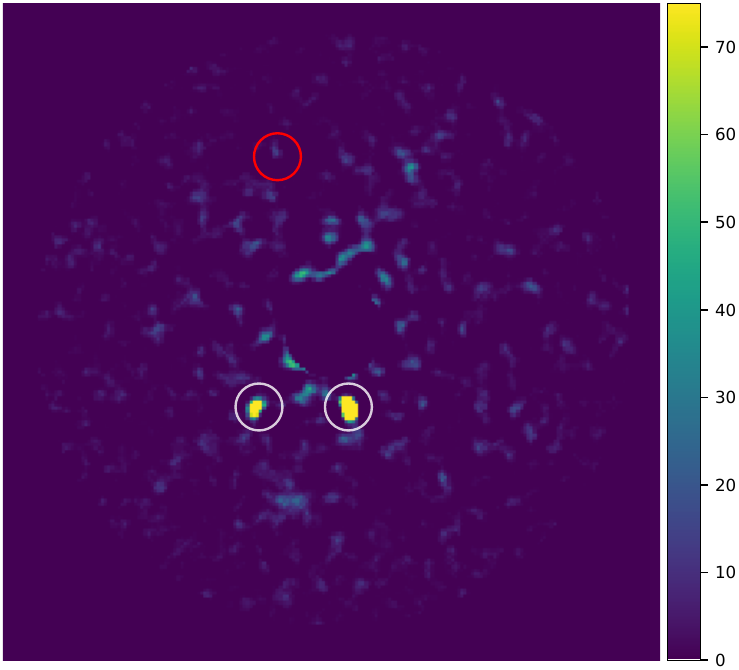}
        \caption{lmr3}
    \end{subfigure}
\caption{LRMs after AMAT$_\textrm{L1}$ for EIDC Datasets: In these maps, white circles represent TP, red square denotes FP, and red circles signify FN.}\label{fig:EIDC_lr}
\end{figure*}

Figure~\ref{fig:EIDC_lr} shows that the version AMAT$_\textrm{L1}$ using LRM proficiently identifies all exoplanets within the VLT/SPHERE-IRDIS datasets and accurately detects four out of seven planets, and six out of seven planets in the LBT/LMIRCam datasets without any false positive in any dataset. These findings showcase the algorithm's high success rate, especially when compared with other algorithms reported in the EIDC results.

\begin{figure}[!t]
    \centering
    \begin{subfigure}[b]
    {\textwidth}
        \includegraphics[width=0.48\textwidth]{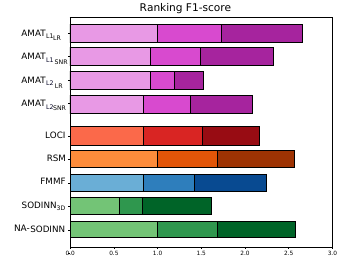}
    \end{subfigure}
\caption{Ranking based on F1-score of the different algorithms in the EIDC results report. The algorithms are classified as AMAT algorithms with S/N maps and LRMs (purple), classical speckle subtraction providing residual maps (red), advanced speckle subtraction building detection maps (orange), inverse problems (blue), and supervised machine learning (green). The light, medium, and dark colors correspond to the three VLT/SPHERE-IRDIS, Keck/NIRC2, and LBT/LMIRCam data sets, respectively.
}\label{fig:EIDC_all_f1}
\end{figure}

In the paper reporting on the EIDC results \citep{cantalloube2020exoplanet}, the algorithms are classified into four categories: classical speckle subtraction providing residual maps, advanced speckle subtraction building detection maps, inverse problems, and supervised machine learning. A comparison of the most successful of these methods in their categories, focusing on the $F1_\textrm{score}$, is presented in Fig.~\ref{fig:EIDC_all_f1}. Our AMAT$_\textrm{L1}$ method with both S/N map and LRM comes out as the most effective within the category of classical speckle subtraction methods. Furthermore, the AMAT$_\textrm{L1}$ method with LRM exhibits a level of success comparable to that of advanced algorithms like RSM and FMMF\@. This underlines the significant potential and robustness of AMAT$_\textrm{L1}$ in the realm of exoplanet detection.

\section{Conclusions}\label{sec:conclusion}
In this paper, we investigated AMAT, a new exoplanet detection method designed to improve the separation of planetary flux from static and quasi-static signals using iterative techniques. The method enhances models based on low-rank approximations like PCA. Current approaches typically assume Gaussian noise, but recent studies suggest that residuals often follow a Laplacian distribution. Most low-rank approximation techniques still rely on PCA, which assumes Gaussian noise. To address this inconsistency, we proposed L1-LRA, which is based on assuming Laplace-distributed noise, and integrated it within the AMAT algorithm.

The AMAT algorithm, which has an open-source implementation, was thoroughly tested with various approaches. First, we compared sensitivity limits using S/N comparisons and contrast curves on the 51 Eri dataset acquired with the VLT/SPHERE-IRDIS instrument. We evaluated the performance of AMAT alongside AnnPCA and AnnL1-LRA, and our results demonstrated significant performance improvements. This was supported by ROC curve comparisons. We then benchmarked AMAT against state-of-the-art algorithms using EIDC datasets. When comparing similar speckle subtraction algorithms, AMAT delivered competitive performance, achieving results comparable to the best in its category. By utilizing the LRM based on Laplacian noise, we further enhanced the algorithm's performance, allowing AMAT$_\textrm{L1}$ with LRM to achieve the highest F1 score among all categories. Additionally, the flux map generated by the AMAT algorithm provided accurate planetary flux measurements, even for faint planets.

One limitation of this study is the high computational cost due to the iterative nature of the algorithm and the need to apply it to each pixel, given the possibility of the planet appearing in any pixel. The computations were performed on a server equipped with an Intel CPU with 18 cores and 125 GB of RAM. While AnnPCA takes only a few minutes to process the 51 Eri dataset, AMAT$_\textrm{L1}$ takes approximately a day to complete the same task. The AMAT$_\textrm{L1}$ algorithm, while producing better results, relies on L1-LRA, which is computationally more demanding than PCA. Future studies should explore faster implementations of L1-LRA to improve the algorithm's speed and applicability. Another limitation involves determining the threshold for the LRM, which should be standardized rather than observation-based, ensuring a more automated and reliable method.

\begin{acknowledgements}
This project has received funding from the European Research Council (ERC) under the European Union’s Horizon 2020 research and innovation programme (grant agreement No 819155); VC thanks the Belgian Federal Science Policy Office (BELSPO) for the provision of financial support in the framework of the PRODEX Programme of the European Space Agency (ESA) under contract number 4000142531; the Fonds de la Recherche Scientifique -- FNRS and the Fonds Wetenschappelijk Onderzoek -- Vlaanderen under EOS Project no 30468160; the Fonds de la Recherche Scientifique-FNRS under Grant no T.0001.23; LJ thanks the FRS-FNRS for the funding related to the PDR project QuadSense (T.0160.24).
\end{acknowledgements}

\bibliographystyle{aa}
\bibliography{references}

\begin{thebibliography}{51}
\expandafter\ifx\csname natexlab\endcsname\relax\def\natexlab#1{#1}\fi

\bibitem[{{Absil} {et~al.}(2013){Absil}, {Milli}, {Mawet}, {Lagrange}, {Girard}, {Chauvin}, {Boccaletti}, {Delacroix}, \& {Surdej}}]{Absil2013}
{Absil}, O., {Milli}, J., {Mawet}, D., {et~al.} 2013, \aap, 559, L12

\bibitem[{Amara \& Quanz(2012)}]{Amara2012pynpoint}
Amara, A. \& Quanz, S.~P. 2012, Monthly Notices of the Royal Astronomical Society, 427, 948

\bibitem[{Beuzit {et~al.}(2019)Beuzit, Vigan, Mouillet, Dohlen, Gratton, Boccaletti, Sauvage, Schmid, Langlois, Petit, {et~al.}}]{beuzit2019sphere}
Beuzit, J.-L., Vigan, A., Mouillet, D., {et~al.} 2019, Astronomy \& Astrophysics, 631, A155

\bibitem[{Bonse {et~al.}(2023)Bonse, Garvin, Gebhard, Dannert, Cantalloube, Cugno, Absil, Hayoz, Milli, Kasper, {et~al.}}]{bonse2023comparing}
Bonse, M.~J., Garvin, E.~O., Gebhard, T.~D., {et~al.} 2023, The Astronomical Journal, 166, 71

\bibitem[{Bowler(2016)}]{bowler2016imaging}
Bowler, B.~P. 2016, Publications of the Astronomical Society of the Pacific, 128, 102001

\bibitem[{Cantalloube {et~al.}(2015)Cantalloube, Mouillet, Mugnier, Milli, Absil, Gonzalez, Chauvin, Beuzit, \& Cornia}]{cantalloube2015direct}
Cantalloube, F., Mouillet, D., Mugnier, L., {et~al.} 2015, Astronomy \& Astrophysics, 582, A89

\bibitem[{Cantalloube {et~al.}(2020)}]{cantalloube2020exoplanet}
Cantalloube, F. {et~al.} 2020, in Adaptive Optics Systems VII, Vol. 11448, SPIE, 1027--1062

\bibitem[{Cantero {et~al.}(2023)Cantero, Absil, Dahlqvist, \& Van~Droogenbroeck}]{cantero2023sodinn}
Cantero, C., Absil, O., Dahlqvist, C.-H., \& Van~Droogenbroeck, M. 2023, arXiv preprint arXiv:2302.02854

\bibitem[{{Christiaens} {et~al.}(2023){Christiaens}, {Gonzalez}, {Farkas}, {Dahlqvist}, {Nasedkin}, {et~al.}}]{VIP_HCI}
{Christiaens}, V., {Gonzalez}, C., {Farkas}, R., {et~al.} 2023, The Journal of Open Source Software, 8, 4774

\bibitem[{Daglayan {et~al.}(2023{\natexlab{a}})Daglayan, Vary, \& Absil}]{daglayan2023esann}
Daglayan, H., Vary, S., \& Absil, P.-A. 2023{\natexlab{a}}, in ESANN 2023-European Symposium on Artificial Neural Networks, Computational Intelligence and Machine Learning

\bibitem[{Daglayan {et~al.}(2022)Daglayan, Vary, Cantalloube, Absil, \& Absil}]{daglayan2022likelihood}
Daglayan, H., Vary, S., Cantalloube, F., Absil, P.-A., \& Absil, O. 2022, in 2022 IEEE 5th International Conference on Image Processing Applications and Systems (IPAS), IEEE, 1--5

\bibitem[{Daglayan {et~al.}(2023{\natexlab{b}})Daglayan, Vary, Leplat, Gillis, \& Absil}]{daglayan2023l1lra}
Daglayan, H., Vary, S., Leplat, V., Gillis, N., \& Absil, P.-A. 2023{\natexlab{b}}, Proceedings of BNAIC/BeNeLearn 2023, 1

\bibitem[{Dahlqvist {et~al.}(2020)Dahlqvist, Cantalloube, \& Absil}]{dahlqvist2020regime}
Dahlqvist, C.-H., Cantalloube, F., \& Absil, O. 2020, Astronomy \& Astrophysics, 633, A95

\bibitem[{Eckart \& Young(1936)}]{eckart1936approximation}
Eckart, C. \& Young, G. 1936, Psychometrika, 1, 211

\bibitem[{Eriksson \& Van Den~Hengel(2010)}]{eriksson2010efficient}
Eriksson, A. \& Van Den~Hengel, A. 2010, in 2010 IEEE Computer Society Conference on Computer Vision and Pattern Recognition, IEEE, 771--778

\bibitem[{Flasseur {et~al.}(2018)Flasseur, Denis, Thi{\'e}baut, \& Langlois}]{flasseur2018exoplanet}
Flasseur, O., Denis, L., Thi{\'e}baut, {\'E}., \& Langlois, M. 2018, Astronomy \& Astrophysics, 618, A138

\bibitem[{Galicher \& Mazoyer(2023)}]{galicher2023imaging}
Galicher, R. \& Mazoyer, J. 2023, Comptes Rendus. Physique, 24, 1

\bibitem[{Gao {et~al.}(2009)Gao, Kwan, \& Guo}]{gao2009robust}
Gao, J., Kwan, P.~W., \& Guo, Y. 2009, Neurocomputing, 72, 1242

\bibitem[{Gillis \& Plemmons(2011)}]{gillis2011dimensionality}
Gillis, N. \& Plemmons, R.~J. 2011, Optical Engineering, 50, 027001

\bibitem[{Gillis \& Vavasis(2018)}]{gillis2018complexity}
Gillis, N. \& Vavasis, S.~A. 2018, Mathematics of Operations Research, 43, 1072

\bibitem[{Gomez~Gonzalez {et~al.}(2016)}]{GomezGonzalez2016Lowranka}
Gomez~Gonzalez, C.~A. {et~al.} 2016, Astronomy \& Astrophysics, 589, A54

\bibitem[{Gomez~Gonzalez {et~al.}(2017)}]{GomezGonzalez2017VIP}
Gomez~Gonzalez, C.~A. {et~al.} 2017, The Astronomical Journal, 154, 7

\bibitem[{Gonzalez {et~al.}(2018)Gonzalez, Absil, \& Van~Droogenbroeck}]{gonzalez2018supervised}
Gonzalez, C.~G., Absil, O., \& Van~Droogenbroeck, M. 2018, Astronomy \& Astrophysics, 613, A71

\bibitem[{Guyon(2018)}]{guyon2018extreme}
Guyon, O. 2018, Annual Review of Astronomy and Astrophysics, 56, 315

\bibitem[{Halko {et~al.}(2011)Halko, Martinsson, \& Tropp}]{halko2011finding}
Halko, N., Martinsson, P.-G., \& Tropp, J.~A. 2011, SIAM review, 53, 217

\bibitem[{Juillard {et~al.}(2023)Juillard, Christiaens, \& Absil}]{juillard2023inverse}
Juillard, S., Christiaens, V., \& Absil, O. 2023, Astronomy \& Astrophysics, 679, A52

\bibitem[{Ke \& Kanade(2003)}]{ke2003robust}
Ke, Q. \& Kanade, T. 2003

\bibitem[{Ke \& Kanade(2005{\natexlab{a}})}]{ke2005alternating}
Ke, Q. \& Kanade, T. 2005{\natexlab{a}}, in 2005 IEEE Computer Society Conference on Computer Vision and Pattern Recognition (CVPR'05), Vol.~1, 739--746 vol. 1

\bibitem[{Ke \& Kanade(2005{\natexlab{b}})}]{ke2005robust}
Ke, Q. \& Kanade, T. 2005{\natexlab{b}}, in 2005 IEEE Computer Society Conference on Computer Vision and Pattern Recognition (CVPR'05), Vol.~1, IEEE, 739--746

\bibitem[{Lafreniere {et~al.}(2007)Lafreniere, Marois, Doyon, Nadeau, \& Artigau}]{lafreniere2007new}
Lafreniere, D., Marois, C., Doyon, R., Nadeau, D., \& Artigau, {\'E}. 2007, The Astrophysical Journal, 660, 770

\bibitem[{Lagrange {et~al.}(2010)Lagrange, Bonnefoy, Chauvin, Apai, Ehrenreich, Boccaletti, Gratadour, Rouan, Mouillet, Lacour, {et~al.}}]{lagrange2010giant}
Lagrange, A.-M., Bonnefoy, M., Chauvin, G., {et~al.} 2010, Science, 329, 57

\bibitem[{Macintosh {et~al.}(2014)Macintosh, Graham, Ingraham, Konopacky, Marois, Perrin, Poyneer, Bauman, Barman, Burrows, {et~al.}}]{macintosh2014first}
Macintosh, B., Graham, J.~R., Ingraham, P., {et~al.} 2014, proceedings of the National Academy of Sciences, 111, 12661

\bibitem[{Marois {et~al.}(2006)Marois, Lafreniere, Doyon, Macintosh, \& Nadeau}]{marois2006angular}
Marois, C., Lafreniere, D., Doyon, R., Macintosh, B., \& Nadeau, D. 2006, The Astrophysical Journal, 641, 556

\bibitem[{Marois {et~al.}(2010)Marois, Macintosh, \& V{\'e}ran}]{marois2010exoplanet}
Marois, C., Macintosh, B., \& V{\'e}ran, J.-P. 2010, in Adaptive Optics Systems II, Vol. 7736, SPIE, 595--606

\bibitem[{Martinache {et~al.}(2009)Martinache, Guyon, Lozi, Garrel, Blain, \& Sivo}]{martinache2009subaru}
Martinache, F., Guyon, O., Lozi, J., {et~al.} 2009, The Subaru coronagraphic extreme AO project

\bibitem[{Mawet {et~al.}(2014)Mawet, Milli, Wahhaj, Pelat, Absil, Delacroix, Boccaletti, Kasper, Kenworthy, Marois, {et~al.}}]{mawet2014fundamental}
Mawet, D., Milli, J., Wahhaj, Z., {et~al.} 2014, The Astrophysical Journal, 792, 97

\bibitem[{{NASA Exoplanet Archive}(2024)}]{nasa}
{NASA Exoplanet Archive}. 2024, Planetary Systems

\bibitem[{Pairet {et~al.}(2019)Pairet, Cantalloube, Gomez~Gonzalez, Absil, \& Jacques}]{pairet2019stim}
Pairet, B., Cantalloube, F., Gomez~Gonzalez, C.~A., Absil, O., \& Jacques, L. 2019, Monthly Notices of the Royal Astronomical Society, 487, 2262

\bibitem[{Pairet {et~al.}(2021)Pairet, Cantalloube, \& Jacques}]{pairet2021mayonnaise}
Pairet, B., Cantalloube, F., \& Jacques, L. 2021, Monthly Notices of the Royal Astronomical Society, 503, 3724

\bibitem[{Ren {et~al.}(2018)Ren, Pueyo, Zhu, Debes, \& Duch{\^e}ne}]{ren2018non}
Ren, B., Pueyo, L., Zhu, G.~B., Debes, J., \& Duch{\^e}ne, G. 2018, The Astrophysical Journal, 852, 104

\bibitem[{Ruffio {et~al.}(2017)Ruffio, Macintosh, Wang, Pueyo, Nielsen, De~Rosa, Czekala, Marley, Arriaga, Bailey, {et~al.}}]{ruffio2017improving}
Ruffio, J.-B., Macintosh, B., Wang, J.~J., {et~al.} 2017, The Astrophysical Journal, 842, 14

\bibitem[{{Samland} {et~al.}(2017){Samland}, {Molli{\`e}re}, {Bonnefoy}, {Maire}, {Cantalloube}, {Cheetham}, {Mesa}, {Gratton}, {Biller}, {Wahhaj}, {Bouwman}, {Brandner}, {Melnick}, {Carson}, {Janson}, {Henning}, {Homeier}, {Mordasini}, {Langlois}, {Quanz}, {van Boekel}, {Zurlo}, {Schlieder}, {Avenhaus}, {Beuzit}, {Boccaletti}, {Bonavita}, {Chauvin}, {Claudi}, {Cudel}, {Desidera}, {Feldt}, {Fusco}, {Galicher}, {Kopytova}, {Lagrange}, {Le Coroller}, {Martinez}, {Moeller-Nilsson}, {Mouillet}, {Mugnier}, {Perrot}, {Sevin}, {Sissa}, {Vigan}, \& {Weber}}]{samland_spectral_2017}
{Samland}, M., {Molli{\`e}re}, P., {Bonnefoy}, M., {et~al.} 2017, \aap, 603, A57

\bibitem[{Song {et~al.}(2017)Song, Woodruff, \& Zhong}]{song2017low}
Song, Z., Woodruff, D.~P., \& Zhong, P. 2017, in Proceedings of the 49th Annual ACM SIGACT Symposium on Theory of Computing, 688--701

\bibitem[{Soummer {et~al.}(2012)Soummer, Pueyo, \& Larkin}]{Soummer2012Detection}
Soummer, R., Pueyo, L., \& Larkin, J. 2012, The Astrophysical Journal, 755, L28

\bibitem[{{Stapper, L. M.} \& {Ginski, C.}(2022)}]{refId0}
{Stapper, L. M.} \& {Ginski, C.} 2022, A\&A, 668, A50

\bibitem[{Thompson \& Marois(2021)}]{thompson2021improved}
Thompson, W. \& Marois, C. 2021, The Astronomical Journal, 161, 236

\bibitem[{Vary {et~al.}(2023)Vary, Daglayan, Jacques, \& Absil}]{vary2023low}
Vary, S., Daglayan, H., Jacques, L., \& Absil, P.-A. 2023, in ICASSP 2023-2023 IEEE International Conference on Acoustics, Speech and Signal Processing (ICASSP), IEEE, 1--5

\bibitem[{Wahhaj {et~al.}(2015)Wahhaj, Cieza, Mawet, Yang, Canovas, De~Boer, Casassus, M{\'e}nard, Schreiber, Liu, {et~al.}}]{wahhaj2015improving}
Wahhaj, Z., Cieza, L.~A., Mawet, D., {et~al.} 2015, Astronomy \& Astrophysics, 581, A24

\bibitem[{Wang {et~al.}(2016)Wang, Graham, Pueyo, Kalas, Millar-Blanchaer, Ruffio, De~Rosa, Ammons, Arriaga, Bailey, {et~al.}}]{wang2016orbit}
Wang, J.~J., Graham, J.~R., Pueyo, L., {et~al.} 2016, The Astronomical Journal, 152, 97

\bibitem[{Wertz {et~al.}(2017)Wertz, Absil, Gonz{\'a}lez, Milli, Girard, Mawet, \& Pueyo}]{wertz2017vlt}
Wertz, O., Absil, O., Gonz{\'a}lez, C.~G., {et~al.} 2017, Astronomy \& Astrophysics, 598, A83

\bibitem[{Zheng {et~al.}(2012)Zheng, Liu, Sugimoto, Yan, \& Okutomi}]{zheng2012practical}
Zheng, Y., Liu, G., Sugimoto, S., Yan, S., \& Okutomi, M. 2012, in 2012 IEEE Conference on Computer Vision and Pattern Recognition, IEEE, 1410--1417

\end{thebibliography}

\appendix

\section{F1-score vs threshold for AMAT$_\textrm{L1}$}

\begin{figure*}[!t]
    \centering
    \begin{subfigure}[b]
    {\textwidth}
        \includegraphics[width=\textwidth]{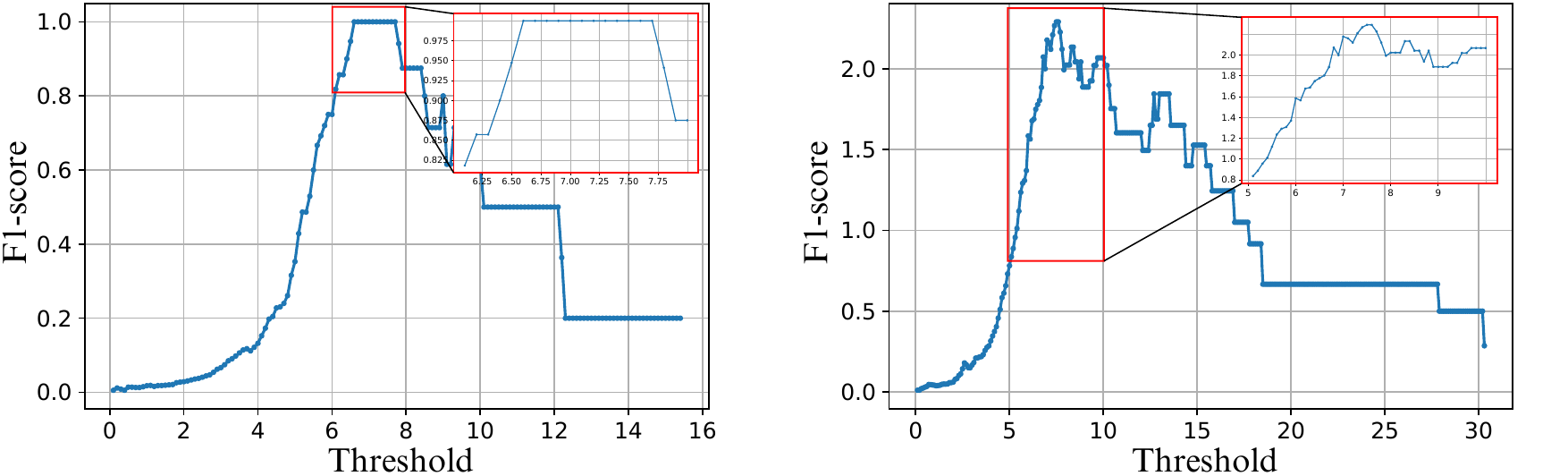}
    \end{subfigure}
\caption{F1-score of AMAT$_{\textrm{L1}}$ algorithm with S/N map using various thresholds for 51 Eri datasets (left) and EIDC datasets (right). }\label{fig:f1_threshold_snr}
\end{figure*}

\begin{figure*}[!t]
    \centering
    \begin{subfigure}[b]
    {\textwidth}
        \includegraphics[width=\textwidth]{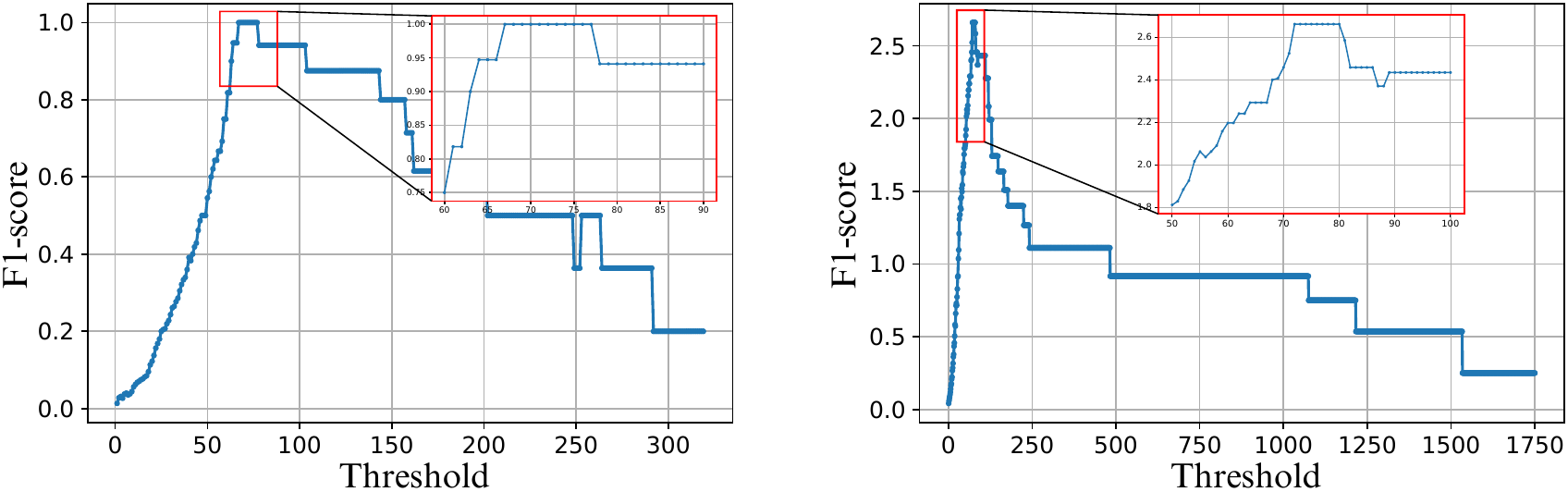}
    \end{subfigure}
\caption{F1-score of AMAT$_{\textrm{L1}}$ algorithm with LRM using various thresholds for 51 Eri datasets (left) and EIDC datasets (right).}\label{fig:f1_threshold}
\end{figure*}

\section{Datasets}
The properties of the datasets which we used in this paper are given in Table~\ref{table:datasets}.

\begin{table}
\caption{The properties of the datasets which we used in the paper. N$_{img}$ is the size of the images, N$_{t}$ is the number of frames, $\Delta_{field}$ is  the total field rotation of the planets, and N$_{p}$ is the number of the injected planet for EIDC datasets and the number of the real planet for 51 Eri dataset.} 
\label{table:datasets}      
\centering                          
\begin{tabular}{c c c c c c}        
\hline\hline 
ID & Telescope/Instr & N$_{img}$ & N$_t$ & $\Delta_{field}$ & N$_p$ \\
& & (px $\times$ px) & & ($^\circ$) & \\ 
\hline                        
   sph1 & VLT/SPHERE-IRDIS & 160$\times$160 & 252 & 40.3 & 1\\      
   sph2 & VLT/SPHERE-IRDIS & 160$\times$160 & 80 & 31.5 & 0\\
   sph3 & VLT/SPHERE-IRDIS & 160$\times$160 & 228 & 80.5 & 5\\
   nrc1 & Keck/NIRC2  & 321$\times$321 & 29 & 53.0& 4\\
   nrc2 & Keck/NIRC2  & 321$\times$321 & 40 & 37.3& 3\\ 
   nrc3 & Keck/NIRC2  & 321$\times$321 &  50  & 166.9 & 0\\
   lmr1 & LBT/LMIRCam & 200$\times$200 & 4838 & 153.4 & 2\\
   lmr2 & LBT/LMIRCam & 200$\times$200 & 3219 & 60.6& 2\\
   lmr3 & LBT/LMIRCam & 200$\times$200 & 4620 & 91.0 & 3\\
   51 Eri & VLT/SPHERE-IRDIS & 200$\times$200 & 256 & 42.0 & 1 \\ 
\hline                                  
\end{tabular}
\end{table}

\section{EIDC results after AnnPCA and after AMAT$_{\textrm{L2}}$}

\begin{figure*}[!t]
    \centering
    \begin{subfigure}[b]
    {0.32\textwidth}
        \includegraphics[width=\textwidth]{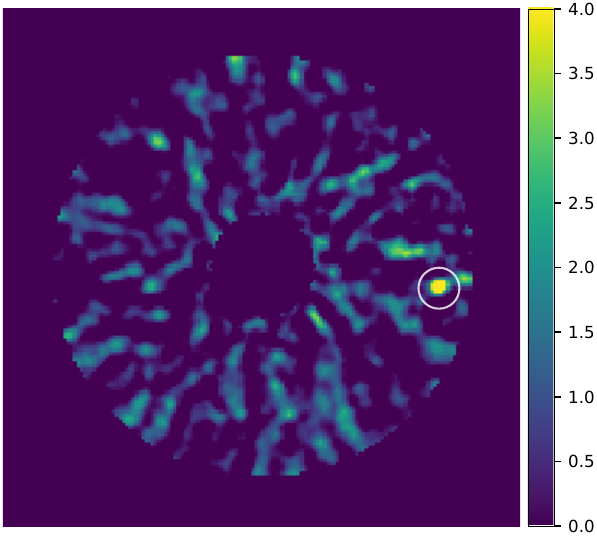}
        \caption{sph1}
    \end{subfigure}
    \begin{subfigure}[b]
    {0.32\textwidth}
        \includegraphics[width=\textwidth]{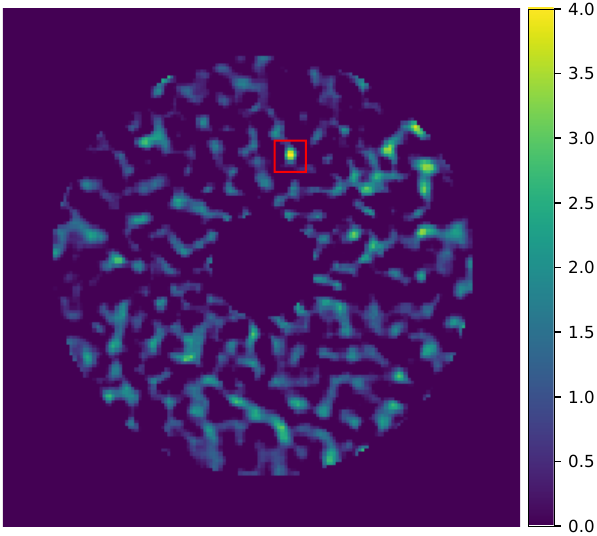}
        \caption{sph2}
    \end{subfigure}
    \begin{subfigure}[b]
    {0.32\textwidth}
        \includegraphics[width=\textwidth]{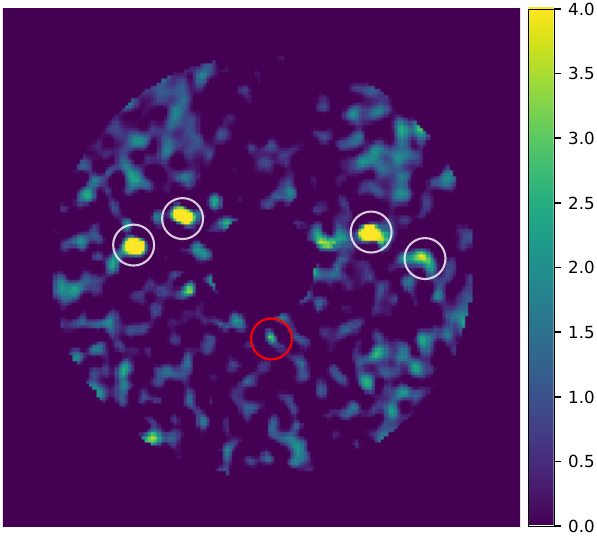}
        \caption{sph3}
    \end{subfigure}
    \begin{subfigure}[b]
    {0.32\textwidth}
        \includegraphics[width=\textwidth]{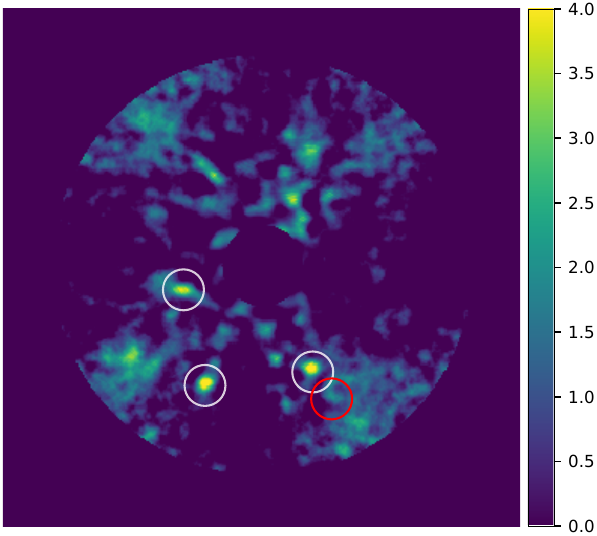}
        \caption{nrc1}
    \end{subfigure}
    \begin{subfigure}[b]
    {0.32\textwidth}
        \includegraphics[width=\textwidth]{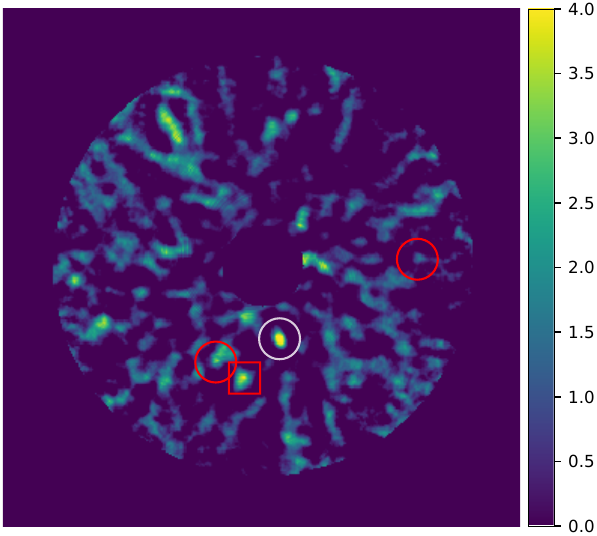}
        \caption{nrc2}
    \end{subfigure}
    \begin{subfigure}[b]
    {0.32\textwidth}
        \includegraphics[width=\textwidth]{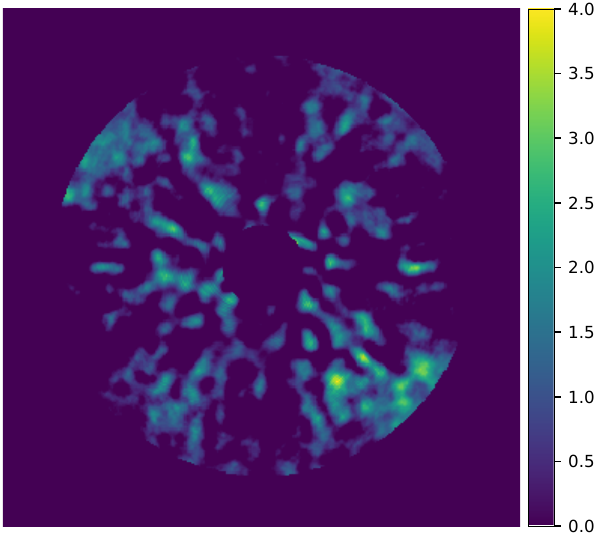}
        \caption{nrc3}
    \end{subfigure}
    \begin{subfigure}[b]
    {0.32\textwidth}
        \includegraphics[width=\textwidth]{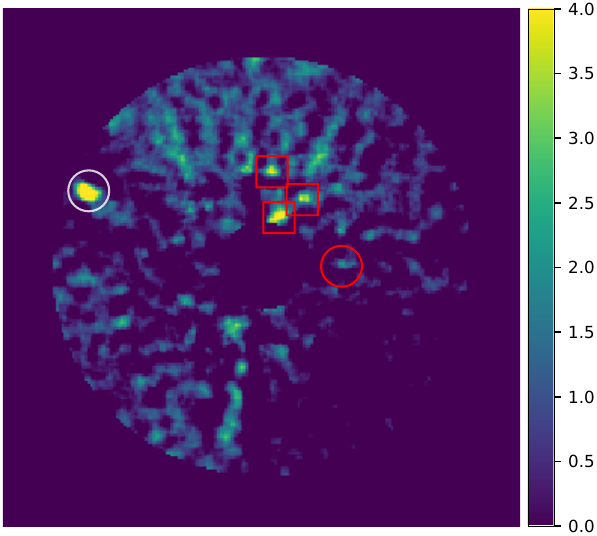}
        \caption{lmr1}
    \end{subfigure}
    \begin{subfigure}[b]
    {0.32\textwidth}
        \includegraphics[width=\textwidth]{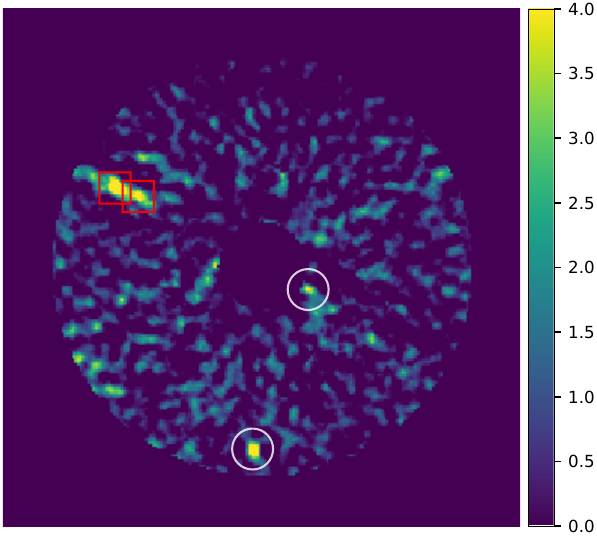}
        \caption{lmr2}
    \end{subfigure}
    \begin{subfigure}[b]
    {0.32\textwidth}
        \includegraphics[width=\textwidth]{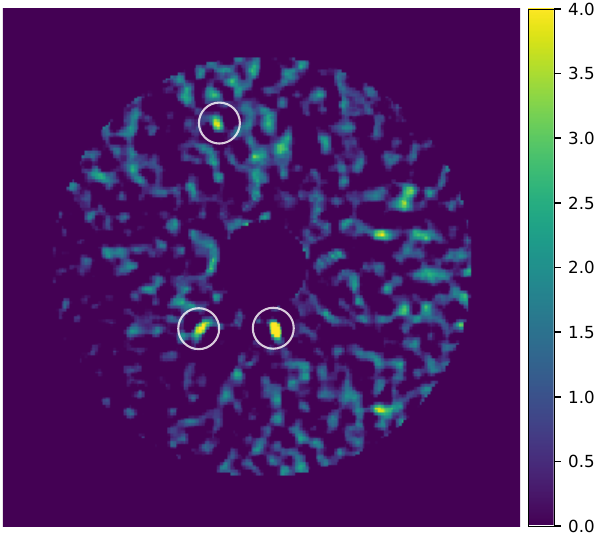}
        \caption{lmr3}
    \end{subfigure}

\caption{S/N maps after AnnPCA using VIP package: In these S/N maps, white circles represent TP, red squares denote FP, and red circles signify FN.}\label{fig:EIDC_snr_apca}
\end{figure*}

\begin{figure*}[!t]
    \centering
    \begin{subfigure}[b]
    {0.32\textwidth}
        \includegraphics[width=\textwidth]{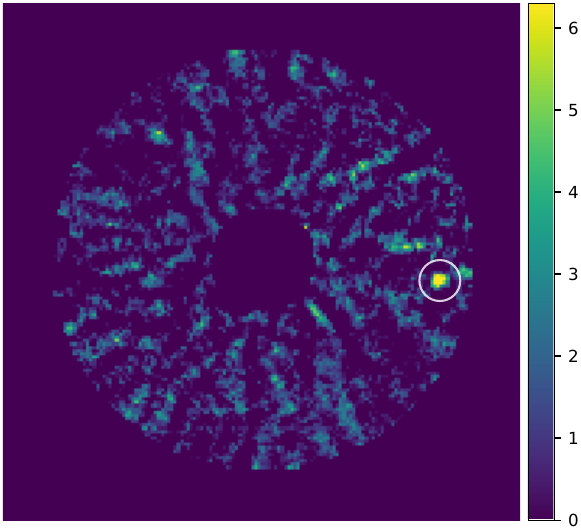}
        \caption{sph1}
    \end{subfigure}
    \begin{subfigure}[b]
    {0.32\textwidth}
        \includegraphics[width=\textwidth]{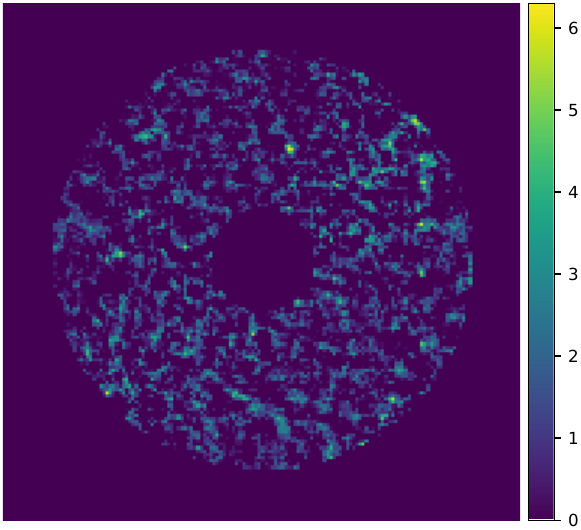}
        \caption{sph2}
    \end{subfigure}
    \begin{subfigure}[b]
    {0.32\textwidth}
        \includegraphics[width=\textwidth]{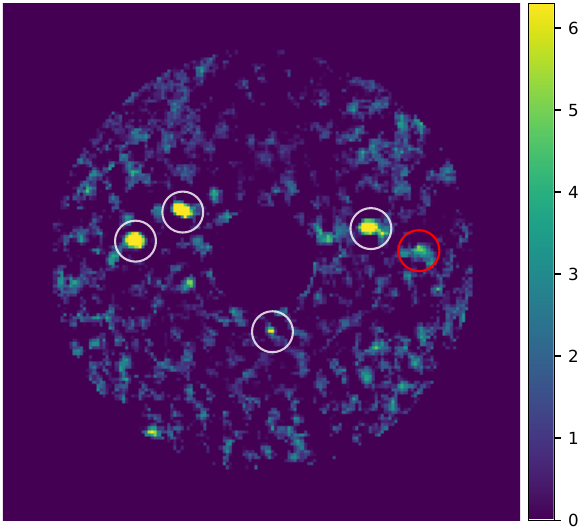}
        \caption{sph3}
    \end{subfigure}
    \begin{subfigure}[b]
    {0.32\textwidth}
        \includegraphics[width=\textwidth]{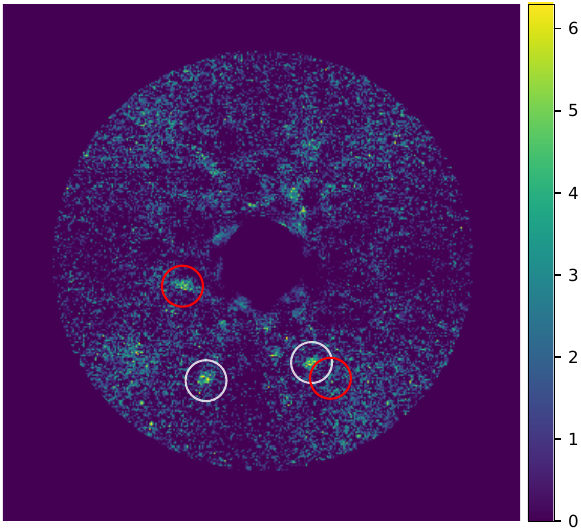}
        \caption{nrc1}
    \end{subfigure}
    \begin{subfigure}[b]
    {0.32\textwidth}
        \includegraphics[width=\textwidth]{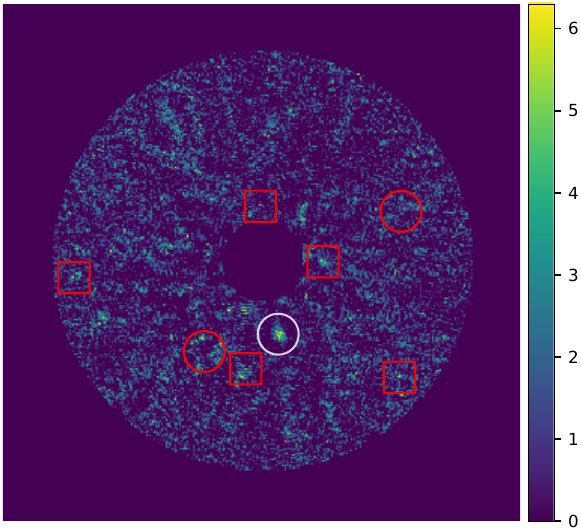}
        \caption{nrc2}
    \end{subfigure}
    \begin{subfigure}[b]
    {0.32\textwidth}
        \includegraphics[width=\textwidth]{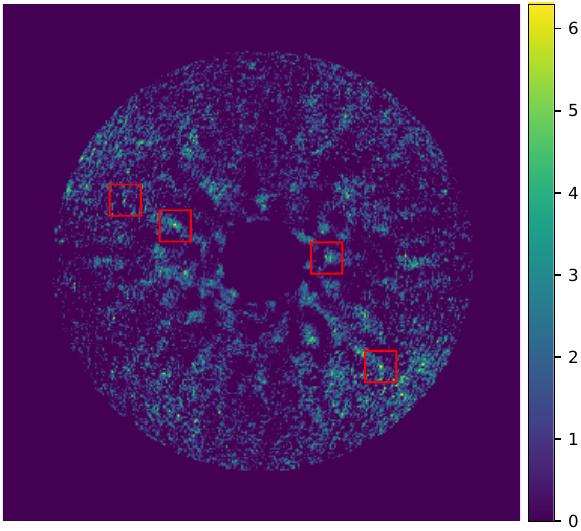}
        \caption{nrc3}
    \end{subfigure}
    \begin{subfigure}[b]
    {0.32\textwidth}
        \includegraphics[width=\textwidth]{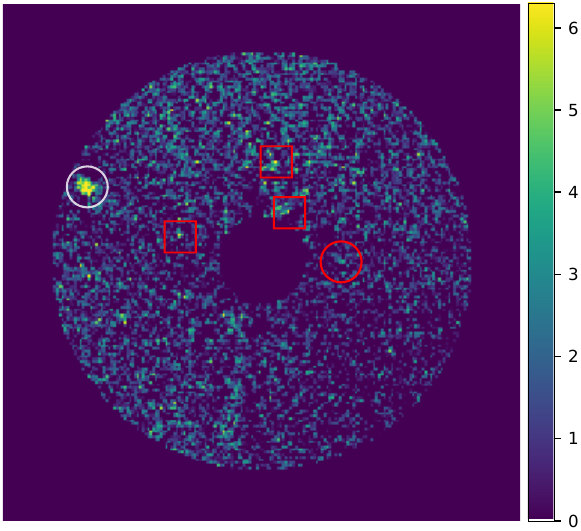}
        \caption{lmr1}
    \end{subfigure}
    \begin{subfigure}[b]
    {0.32\textwidth}
        \includegraphics[width=\textwidth]{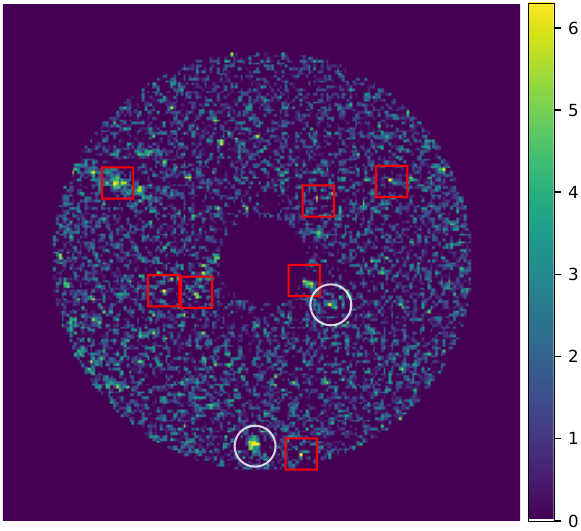}
        \caption{lmr2}
    \end{subfigure}
    \begin{subfigure}[b]
    {0.32\textwidth}
        \includegraphics[width=\textwidth]{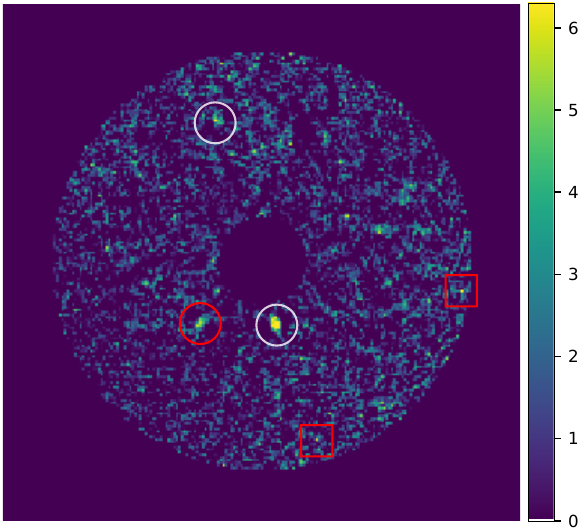}
        \caption{lmr3}
    \end{subfigure}

\caption{S/N maps, proposed by \citet{bonse2023comparing}, after AnnPCA: In these S/N maps, white circles represent TP, red squares denote FP, and red circles signify FN.}\label{fig:EIDC_snr_apca_center}
\end{figure*}

\begin{figure*}[!t]
    \centering
    \begin{subfigure}[b]
    {0.32\textwidth}
        \includegraphics[width=\textwidth]{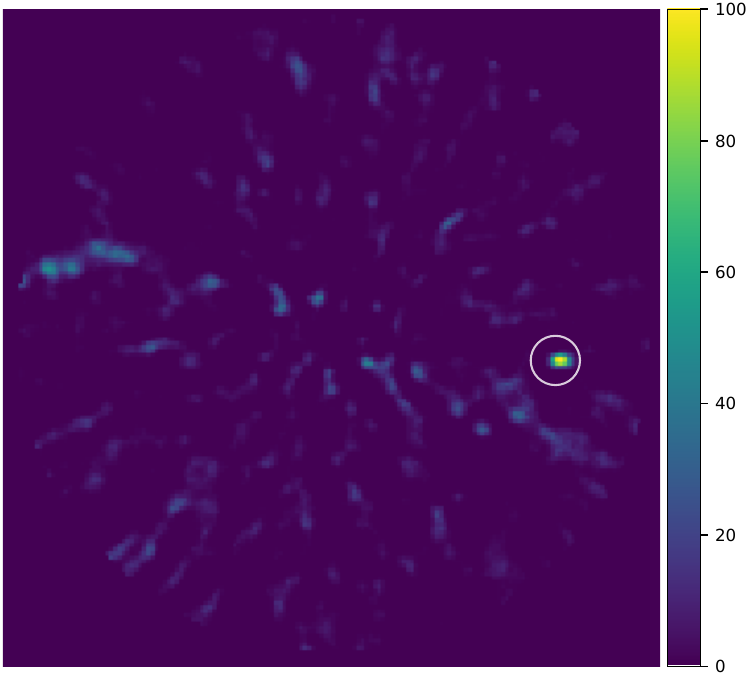}
        \caption{sph1}
    \end{subfigure}
    \begin{subfigure}[b]
    {0.32\textwidth}
        \includegraphics[width=\textwidth]{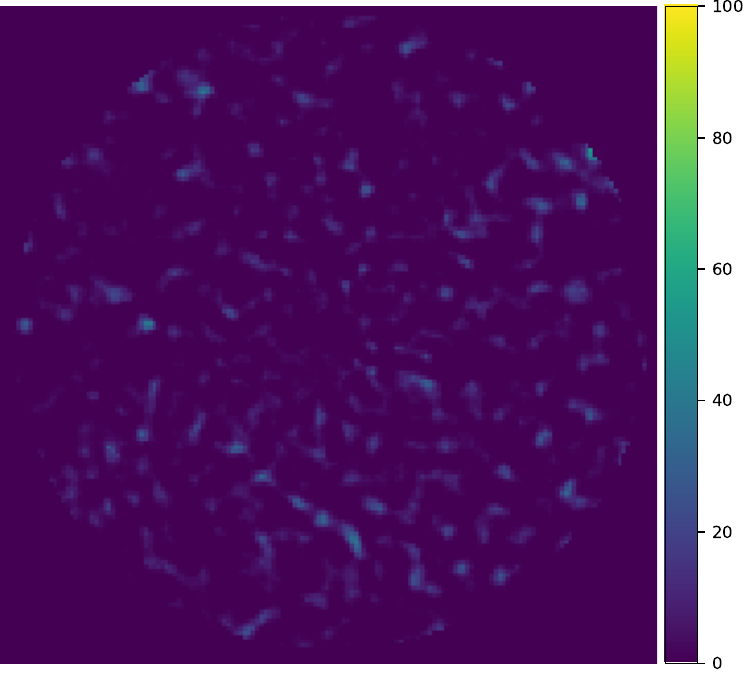}
        \caption{sph2}
    \end{subfigure}
    \begin{subfigure}[b]
    {0.32\textwidth}
        \includegraphics[width=\textwidth]{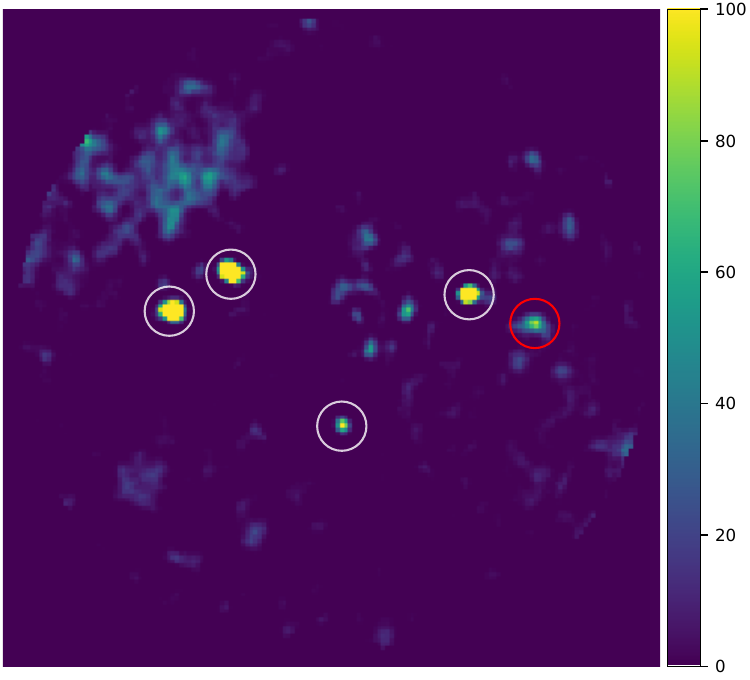}
        \caption{sph3}
    \end{subfigure}
    \begin{subfigure}[b]
    {0.32\textwidth}
        \includegraphics[width=\textwidth]{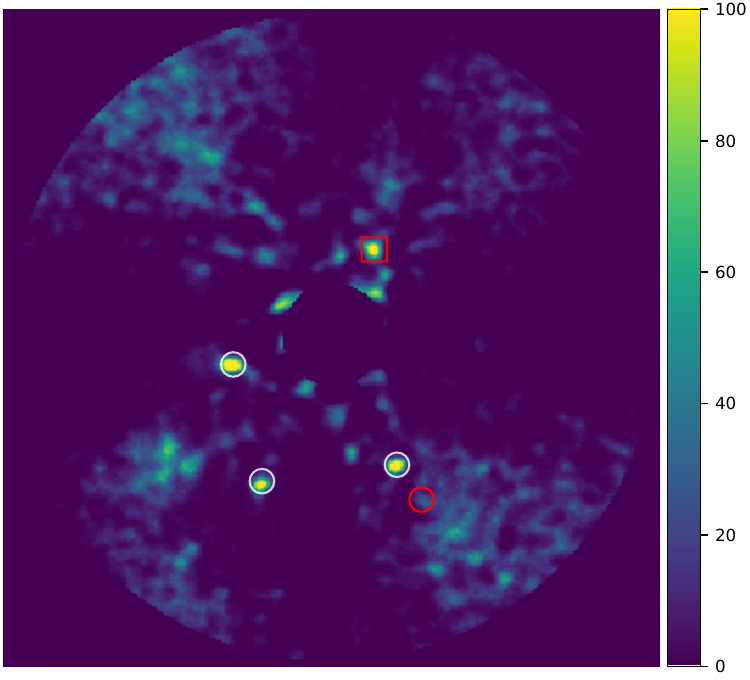}
        \caption{nirc1}
    \end{subfigure}
    \begin{subfigure}[b]
    {0.32\textwidth}
        \includegraphics[width=\textwidth]{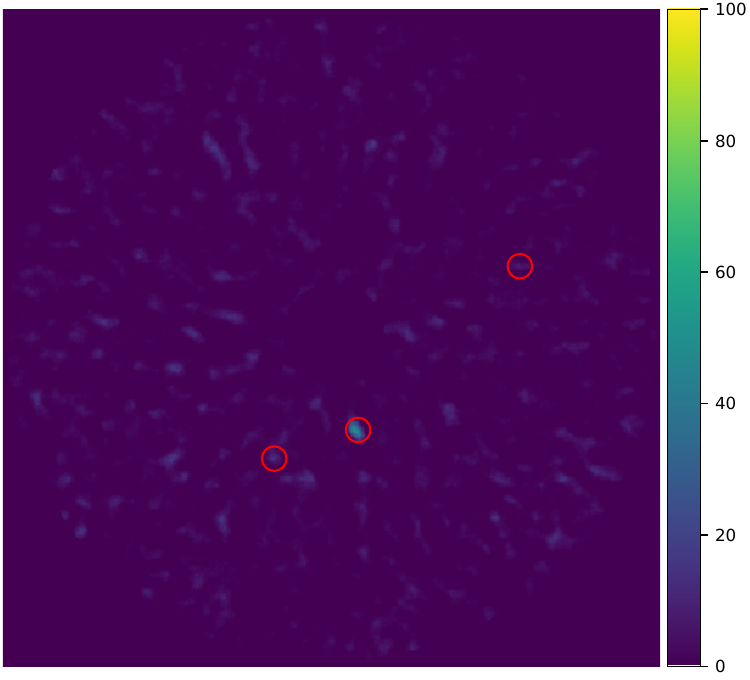}
        \caption{nirc2}
    \end{subfigure}
    \begin{subfigure}[b]
    {0.32\textwidth}
        \includegraphics[width=\textwidth]{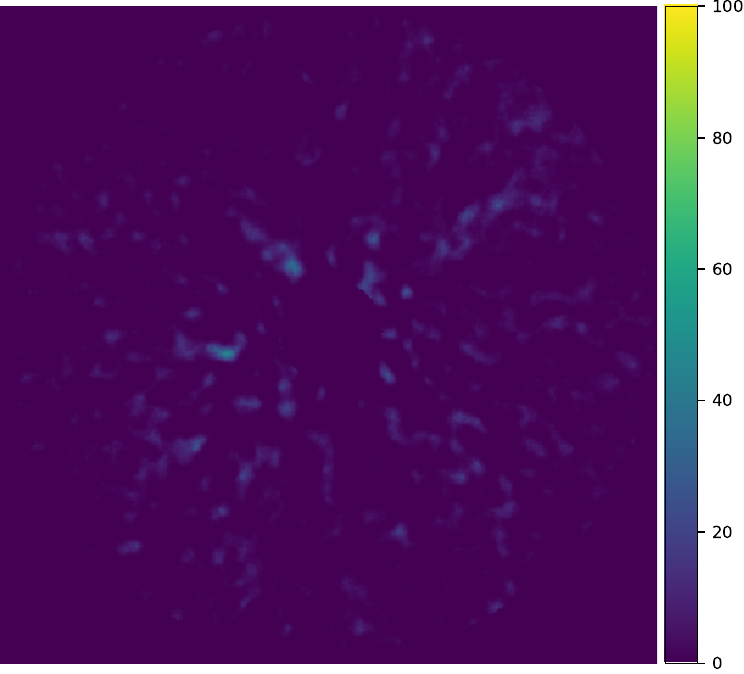}
        \caption{nirc3}
    \end{subfigure}
    \begin{subfigure}[b]
    {0.32\textwidth}
        \includegraphics[width=\textwidth]{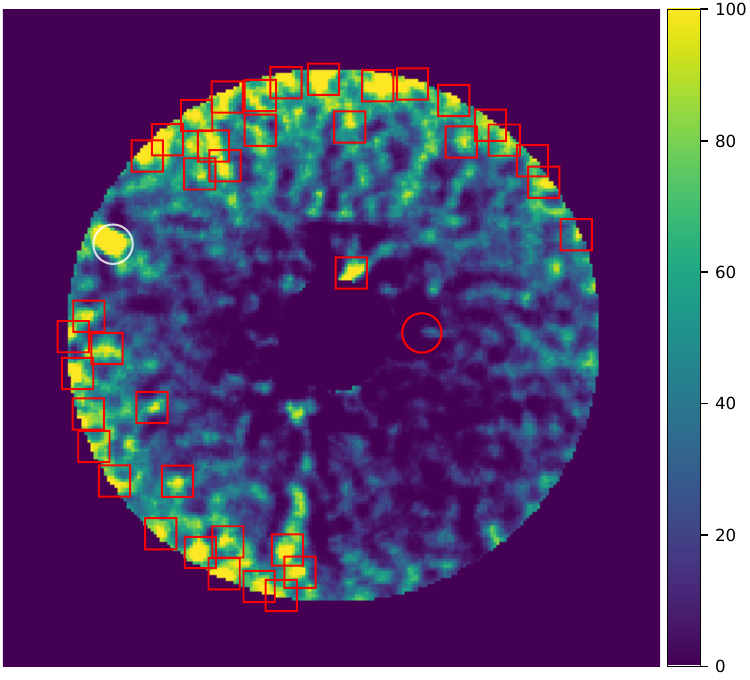}
        \caption{lmr1}
    \end{subfigure}
    \begin{subfigure}[b]
    {0.32\textwidth}
        \includegraphics[width=\textwidth]{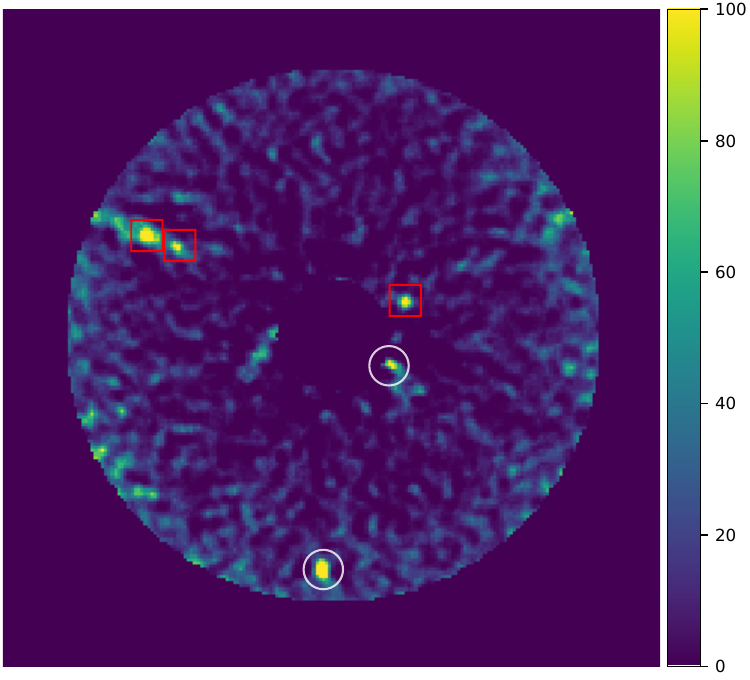}
        \caption{lmr2}
    \end{subfigure}
    \begin{subfigure}[b]
    {0.32\textwidth}
        \includegraphics[width=\textwidth]{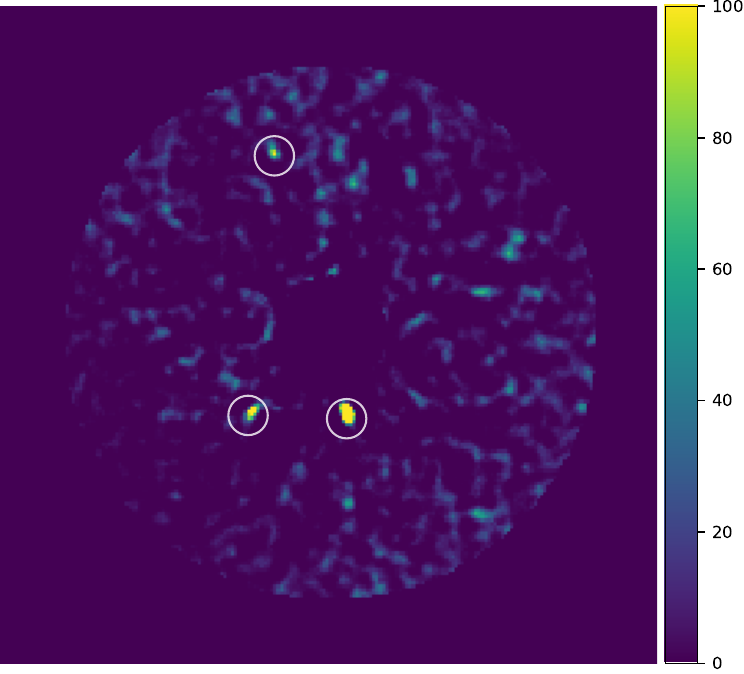}
        \caption{lmr3}
    \end{subfigure}
\caption{LRM after AnnPCA for EIDC Datasets: In these maps, white circles represent TP, red squares denote FP, and red circles signify FN.}\label{fig:EIDC_lr_apca}
\end{figure*}

\section{The location of injected planets in 51 Eri dataset.}\label{sec:eri_fake}

\begin{figure*}[!t]
    \centering
    \begin{subfigure}[b]
    {0.32\textwidth}
        \includegraphics[width=\textwidth]
        {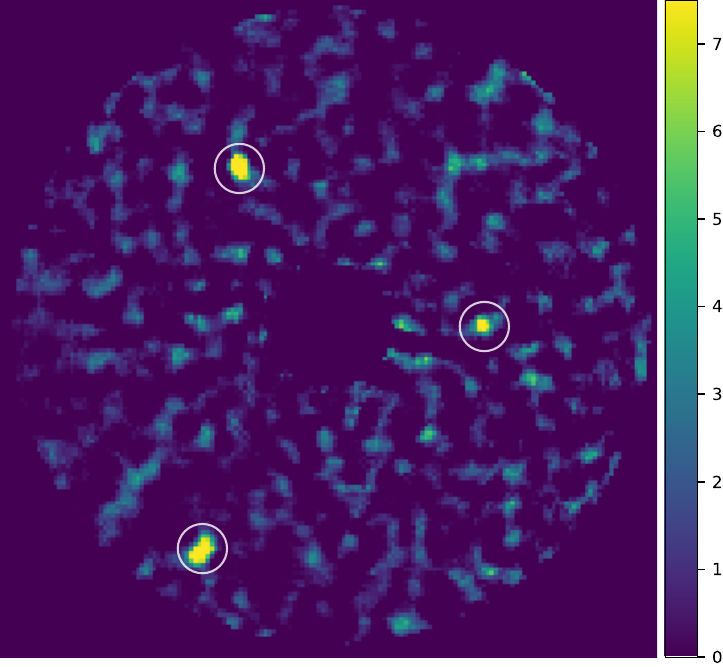}
    \end{subfigure}
    \begin{subfigure}[b]
    {0.32\textwidth}
        \includegraphics[width=\textwidth]{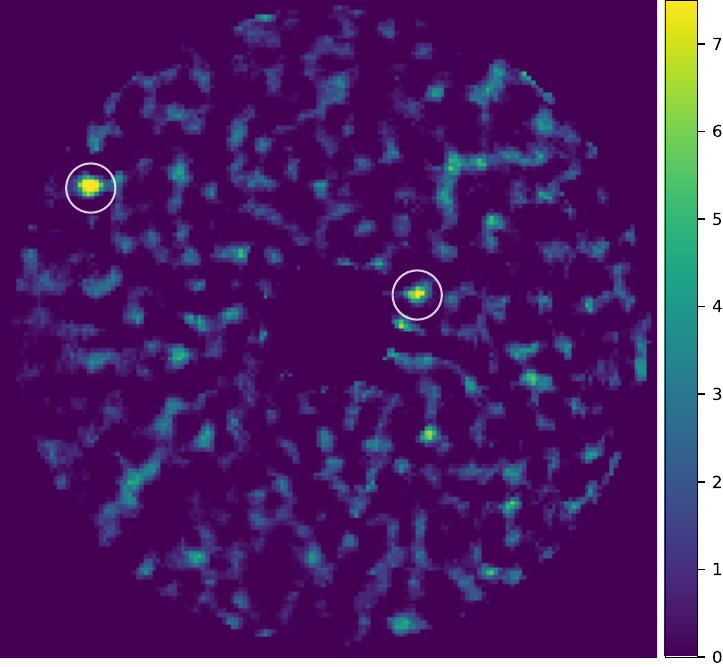}
    \end{subfigure}
    \begin{subfigure}[b]
    {0.32\textwidth}
        \includegraphics[width=\textwidth]{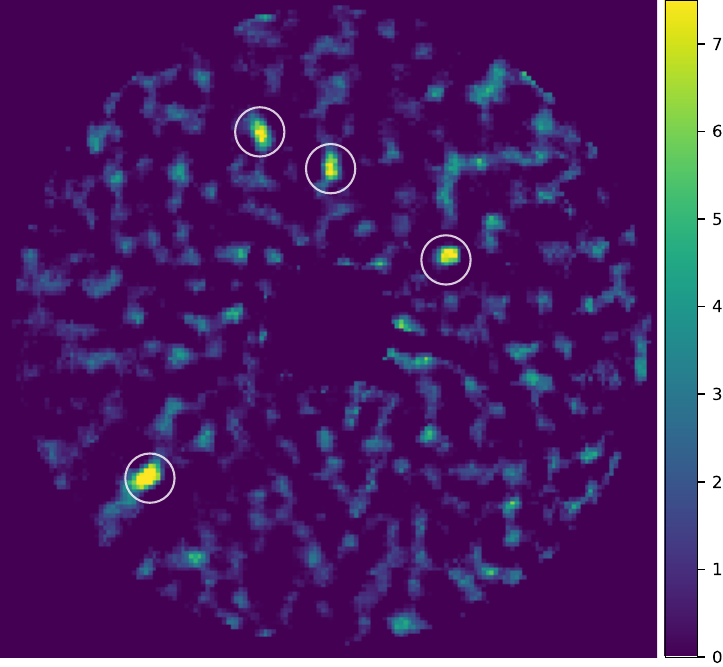}
    \end{subfigure}
\caption{S/N maps for 51 Eri datasets with injected planets: In these S/N maps, white circles represent TP.}\label{fig:Eri51_injecteds_snr}
\end{figure*}

\begin{figure*}[!t]
    \centering
    \begin{subfigure}[b]
    {0.32\textwidth}
        \includegraphics[width=\textwidth]
        {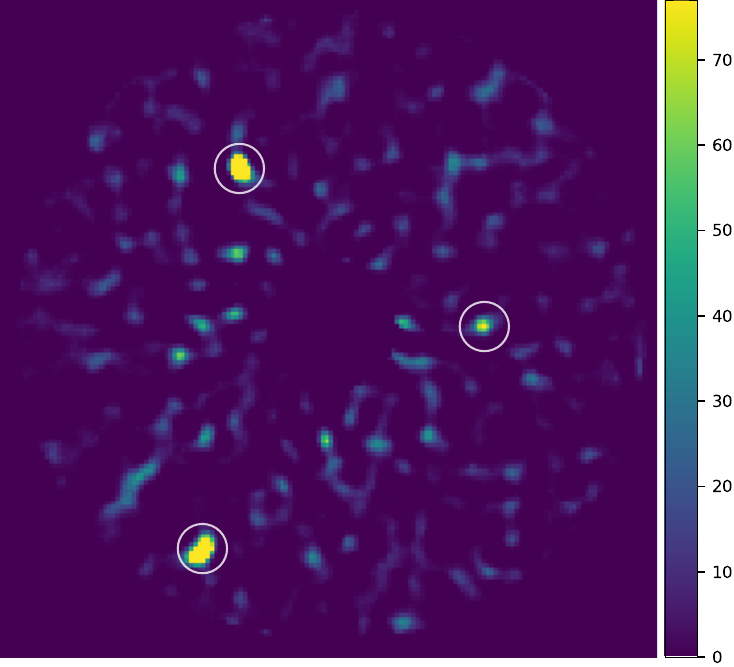}
    \end{subfigure}
    \begin{subfigure}[b]
    {0.32\textwidth}
        \includegraphics[width=\textwidth]{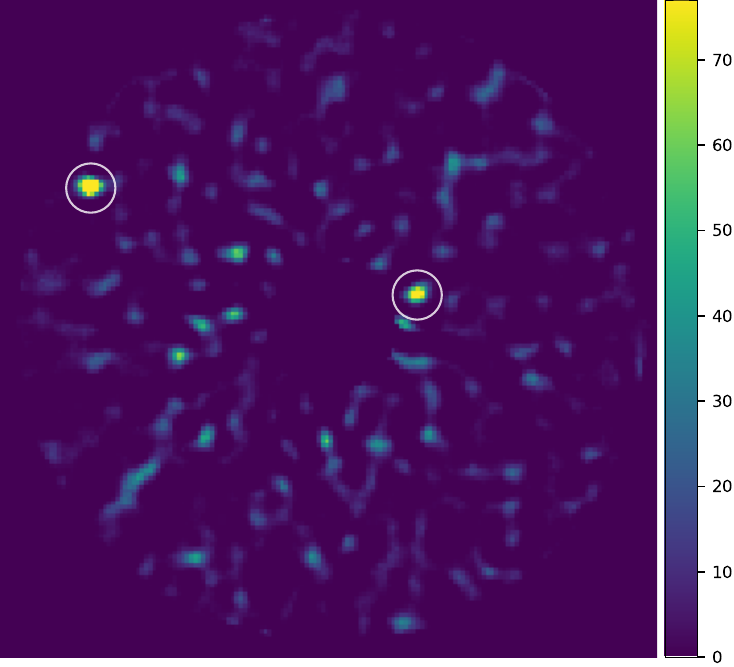}
    \end{subfigure}
    \begin{subfigure}[b]
    {0.32\textwidth}
        \includegraphics[width=\textwidth]{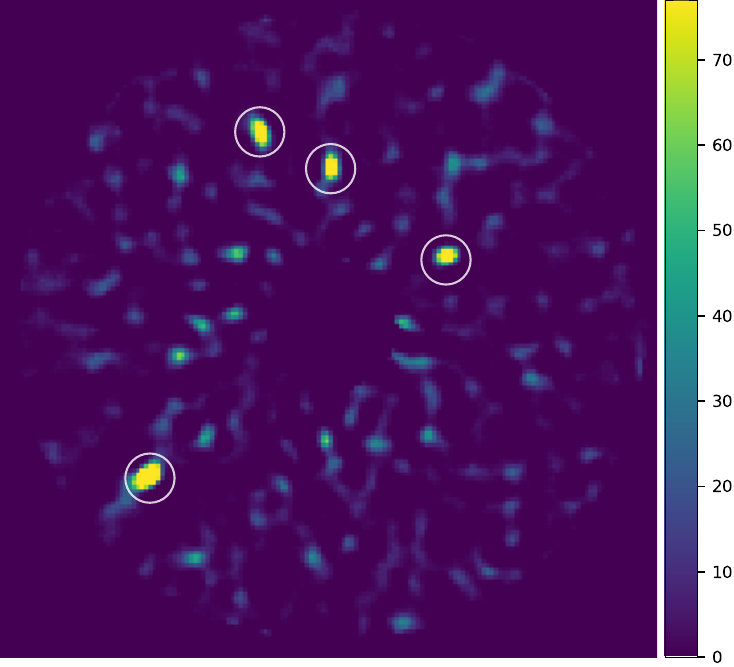}
    \end{subfigure}
\caption{LRMs for 51 Eri datasets with injected planets: In these S/N maps, white circles represent TP.}\label{fig:Eri51_injecteds}
\end{figure*}

S/N maps and LRMs of 51 Eri datasets with injected planets are given in Fig.~\ref{fig:Eri51_injecteds_snr} and \ref{fig:Eri51_injecteds}, respectively.

\end{document}